\documentclass[a4paper, journal]{IEEEtran}
\usepackage[cmex10]{amsmath}
\usepackage{hhline}
\usepackage{amsmath,graphicx,multirow}
\usepackage{times,epsfig}
\usepackage{bm,amssymb}
\usepackage{color, soul}
\usepackage{algorithm,algpseudocode}
\usepackage{eqparbox}
\usepackage{epstopdf}
\usepackage{framed}
\usepackage{array}
\usepackage{mdwmath}
\usepackage{mdwtab}
\usepackage{amscd}
\usepackage[tight,footnotesize]{subfigure}

\begin{document}
%
\title{\Huge{Bernoulli-Gaussian Approximate Message-Passing Algorithm for Compressed Sensing with \\1D-Finite-Difference Sparsity}}

\author{ Jaewook~Kang,~\IEEEmembership{Student  Member,~IEEE,}
        Hyoyoung~Jung,~\IEEEmembership{Student Member,~IEEE,}\\
        Heung-No~Lee,~\IEEEmembership{Senior Member,~IEEE,}
        and~Kiseon~Kim,~\IEEEmembership{Senior Member,~IEEE}
\vspace{-20pt}

 \thanks{ The authors are with Department of Information and Communication, Gwangju Institute of Science and
 Technology, Gwangju, Republic of Korea
 (Email:\{jwkkang,rain,heungno,kskim\}@gist.ac.kr)}

 \thanks{Portions of this work were presented at \emph{48th Asilomar Conf. on Signals, Systems, and Computers} (Pacific Grove, CA), Nov. 2014 \cite{ssAMP1}.}
}

\markboth{IEEE Transactions on Signal Processing Draft, VOL. XX, NO.
XX,  2015}{J.Kang \MakeLowercase{\textit{et al.}}}
 \maketitle

\setlength{\baselineskip}{0.97\baselineskip}
\begin{abstract}
This paper proposes a fast approximate message-passing (AMP)
algorithm for solving compressed sensing (CS) recovery problems with
1D-finite-difference  sparsity in term of MMSE estimation. The
proposed algorithm, named ssAMP-BGFD, is low-computational with its
fast convergence and cheap per-iteration cost, providing phase
transition nearly approaching to the state-of-the-art. The proposed
algorithm is originated from a sum-product message-passing rule,
applying a Bernoulli-Gaussian (BG) prior, seeking an MMSE solution.
The algorithm construction includes not only the conventional AMP
technique for the measurement fidelity, but also suggests a
simplified message-passing method to promote the signal sparsity in
finite-difference. Furthermore, we provide an EM-tuning methodology
to learn the BG prior parameters, suggesting how to use some
practical measurement matrices satisfying the RIP requirement under
the ssAMP-BGFD recovery. Extensive empirical results confirms
performance of the proposed algorithm, in phase transition,
convergence speed, and CPU runtime, compared to the recent
algorithms.
\end{abstract}


\section{Introduction}
\subsection{Background}
We consider  compressed sensing (CS) recovery problems for
estimating piecewise-constant (PWC) signals $\underline X \in
\mathbb{R}^N$, whose sparsity is in its 1D-finite-difference (FD),
from  noisy measurements $\underline Y\in \mathbb{R}^M$ given by
\begin{align}\label{system}
\underline Y = {\mathbf{H}}\underline X + \underline W,
\end{align}
where $\underline W \in \mathbb{R}^M$ is handled as an AWGN vector,
and ${\mathbf{H}} \in {\mathbb{R}^{M \times N}}$ is a  measurement
matrix. In particular, we deal with incomplete measurements
$\underline Y$ such that the linear system \eqref{system} is
underdetermined, meaning that the number of measurements $M$ is
significantly smaller than the signal length $N$ $(M \ll N)$.

The 1D-PWC signal model has been mainly used in bioinformatics or
computational biology applications such as  genomic data analysis
\cite{witten1},\cite{witten2},\cite{fusedlasso} and analysis of
molecular dynamics for bacteria \cite{molecular1},\cite{molecular2}.
For such applications, compressed sensing  can be a promising DSP
technique with its dimensionality reduction and sparsity-based
denoising abilities because the biological signals/data  are
basically noisy, requiring large memory storage for its huge
datasize. We further introduce an excellent work of Little and Jones
discussing various types of the PWC signal model and its
applications  \cite{PWC}.

Since the solution finding of \eqref{system} is ill-posed,
optimization methods with regularization have been mostly
considered. This allows us to pick an unique point ${\widehat
{\underline X }}$ from the solution space by imposing a suitable
regularizer of $\underline X$. The most classical regularizer to the
present problem is \emph{total variation} (TV) \cite{TV}. In the TV
method, the sparsity of $\underline X$ can be promoted by applying
an 1D-FD operator defined as ${\mathbf{D}}\underline X  = {[{X_2} -
{X_1},{X_3} - {X_2},...,{X_N} - {X_{N - 1}}]^T}$. Then, the TV
regularization  is represented as a non-smooth convex optimization
\cite{Chen},\cite{candes1}, given as
\begin{align}\label{TV}
{\widehat {\underline X }} = \arg \mathop {\min
}\limits_{\underline X } \left\| {\underline Y  -
{\mathbf{H}}\underline X } \right\|_2^2 +
{\lambda}||{\mathbf{D}}\underline X |{|_1},
\end{align}
where  the parameter $\lambda>0$ controls the balance between the FD sparsity and measurement fidelity which is measured by the squared-error term $\left\| {\underline Y  -
{\mathbf{H}}\underline X } \right\|_2^2$. In statistics area, the optimization method \eqref{TV} is called \emph{Fused Lasso} \cite{fusedlasso}.

One popular approach for solving \eqref{TV} is to use the
first-order algorithms which provide global convergence  in the
general class of convex optimizations, whose convergence rate can be
accelerated  by applying the Nesterov's method \cite{Nesterov}. As
practical first-order solvers, \emph{Chambolle-Pock} (TV-CP)
\cite{CP}, \emph{Fast Iterative Soft-Thresholding Algorithm} (FISTA)
\cite{FISTA}, and \emph{Efficient Fused Lasso} (EFLA) \cite{sfa}
have been highlighted recently.


\subsection{Contribution}
In the present work, we revisit the CS recovery problem with  the
Bayesian philosophy, recasting the problem to
\begin{align}\label{BMMSE}
\begin{gathered}
  \underline {\widehat X}  = \arg \min_{\underline x} \int {\int {{{(\underline x  - \widehat {\underline x })}^2}{f_{\underline X ,\underline Y }}(\underline x ,\underline y )d\underline x d\underline y } } \hfill \\
   \,\,\,\,\,\,\,\,= \frac{1}{Z}\int {\underline x\,\underbrace{{f_{\underline X }}(\mathbf{D}\underline x)}_{\text{Prior}}\underbrace{\mathcal{N}(\mathbf{H}\underline x;\underline y ,\Delta)}_{\text{Likelihood}} d\underline x } \hfill \\
   \end{gathered}
\end{align}
by applying \emph{minimum-mean-square-error} (MMSE) estimation
method \cite{MAP_MMSE}, where $Z>0$ is a normalization constant
independent of $\underline x$. The main advantage of the MMSE method
over the TV method \eqref{TV} is  the MMSE-optimality. It guarantees
better reconstruction quality in terms of MSE if the signal
statistics has a good match with the given prior ${f_{\underline X
}}(\cdot)$ \cite{GAMP},\cite{EM-BG-GAMP},\cite{MAP_MMSE}. On the
other hand, one critical disadvantage is analytical intractability
of the integral calculation of \eqref{BMMSE}.

The main focus of this paper is on low-computational solving of the
CS recovery with the 1D-FD sparsity, and for this purpose we
approach the MMSE estimation \eqref{BMMSE} using
 Bayesian \emph{approximate message-passing} (AMP) which is an approximate loopy belief propagation (BP) based on the central-limit-theorem (CLT) \cite{AMP1}-\cite{EM-BP}.
This is motivated by the fact that the ``sum-product" mode of the
Bayesian AMP provides  accurate and low-computational approximation
to the posterior information for \eqref{BMMSE}, which have been
demonstrated in the CS recovery with direct sparsity
\cite{GAMP}-\cite{EM-BP}. The AMP approach also has shown their
usefulness by providing own \emph{mean-square-error} (MSE)
prediction method, called \emph{state evolution}\footnote{ The state
evolution method is confined to cases with
 \emph{i.i.d.}-random $\mathbf{H}$ and  the signal estimation function \eqref{BMMSE} which  is  scalar-separable and
Lipschitz-continuous \cite{Montanari},\cite{Bayati},\cite{Bayati2}.}
\cite{AMP1},\cite{Montanari},\cite{Bayati}.

To the MMSE method \eqref{BMMSE}, we propose a Bayesian AMP
algorithm using a Bernoulli-Gaussian (BG) prior. We adopt the BG
prior as a key to resolve the analytical intractability of the MMSE
method. The proposed AMP is referred to as \emph{Spike-and-Slab
Approximate Message-Passing using Bernoulli-Gaussian
finite-difference prior} (ssAMP-BGFD), which was partially
introduced in our short paper \cite{ssAMP1}\footnote{The proposed
algorithm was previously named as ``ssAMP-1D" in our conference
paper \cite{ssAMP1}.}. We claim that ssAMP-BGFD has advantages  in
the present CS reconstruction problem, as following:
\begin{itemize}
\item  ssAMP-BGFD  shows  phase transition (PT) closely approaching  to the state-of-the-art performance.
\item ssAMP-BGFD  provides low-computationality which is originated from its cheap per-iteration cost with $\mathcal{O}(MN)$ and  its fast
convergence characteristic.
\item ssAMP-BGFD is compatible with  several non-\emph{i.i.d.}-random matrices $\mathbf{H}$.
\item ssAMP-BGFD optionally provides \emph{Expectation-Maximization} (EM) tuning for prior parameters. Therefore,
ssAMP-BGFD can be parameter-free.
\end{itemize}
Each statement claimed above will be discussed and validated in the
main body of this paper.

TVAMP, proposed by Donoho \emph{et al.} in  \cite{TV_AMP}, can be
considered as another AMP option for the present problem, which is
an extension of the standard AMP \cite{AMP1} applying an anisotropic
TV denoising to estimate $\underline X$. Therefore, TVAMP has a
simple algorithmic structure, providing highly fast solution to the
problem. However, our empirical result reveals that its PT
characteristic is apart from the state-of-the-art.

While working on ssAMP-BGFD, we became aware of an relevant AMP work
by Schniter \emph{et al.}, named GrAMPA \cite{GrAMPA}, which was
carried out independently and concurrently with our work. The GrAMPA
algorithm is based on a novel configuration of the generalized AMP
(GAMP) package \cite{GAMP} for the analysis CS setup
\cite{analysisCS}, providing the both Bayesian options: MMSE and MAP
estimation. Therefore, it is not confined to this CS problem with
1D-FD sparsity, but being applicable to the problem with generalized
sparsity. In addition, it has been empirically confirmed that GrAMPA
shows the state-of-the-art PT performance in the CS recovery with
1D-FD sparsity \cite{ssAMP1}.

We argue that  ssAMP-BGFD is practically advantageous over TVAMP and
GrAMPA for the present problem. ssAMP-BGFD provides its solution
$\widehat{\underline X}$ as simple and fast as TVAMP does, while
showing PT characteristic nearly approaching to the state-of-the-art
by GrAMPA. In addition, TVAMP and GrAMPA require  parameter
configuration before running it, whereas ssAMP-BGFD does not with an
auto-parameter tuning by an Expectation Maximization (EM) technique.
Furthermore, we empirically demonstrate that
column-sign-randomization \cite{candes},\cite{krahmer} enables
ssAMP-BGFD to work well with practical non-\emph{i.i.d.}-random
matrices satisfying the RIP requirement: sub-sampled \emph{Discrete
Cosine} and \emph{Walsh-Hadamard} Transforms, quasi-Toeplitz, and
deterministic bipolar (proposed in \cite{Amini}) matrices. We also
check the compatibility of ssAMP-BGFD with random sparse matrices.

\subsection{Organization and Notation}
The remainder of the paper is organized as follows. Section II is
devoted for a brief introduction to the AMP fundamental and the two
existing AMP algorithms for the CS recovery with 1D-FD sparsity:
TVAMP \cite{TV_AMP} and GrAMPA \cite{GrAMPA}. Section III describes
the construction details of the proposed algorithm, ssAMP-BGFD. In
Section IV, we provides extensive empirical results to validate
several aspects of the ssAMP-BGFD algorithm, compared to the two
AMP-based solvers, TVAMP and GrAMPA, as well as the two first-order
solvers for the TV method \eqref{TV}, \emph{Efficient Fused Lasso}
 \cite{sfa} and \emph{Chambolle-Pock}  \cite{CP}. In Section
V, we provide a practical example of the ssAMP-BGFD  recovery to a
genomic data set. Finally, we conclude this paper in Section VI.

Throughout the paper, we use the following notation. We use
underlined letter like $\underline h$ to denote vectors, boldface
capital letters like $\mathbf{H}$ to denote matrices, and
calligraphic capital letters like $\mathcal{F}$ to indicate set
symbols. The vectors $\underline 1 \equiv [1,...,1]^T$ and
$\underline 0 \equiv [0,...,0]^T$ denote an one vector and a zero
vector respectively. In addition, $f_{X_i}(x_i)$ is a probability
density function (PDF) of a random variable  $X_i \sim f_{X_i}(x_i)$
and its realization is denoted by small letters like $x_i$. We use
$\mathbb{E}_{f_{X_i}}[\cdot]$, $\mathbf{Var}_{f_{X_i}}[\cdot]$, and
$\mathbb{H}({f_{X_i}})$ to denote the expectation, the variance, and
the information entropy with respect to the PDF ${f_{X_i}(x)}$,
respectively. For PDF notation, we use $\mathcal{N}(x_i;\mu ,{\sigma
^2}) \equiv \frac{1}{{\sqrt {2\pi {\sigma ^2}} }}\exp \left( { -
\frac{{{{(x_i - \mu )}^2}}}{{2{\sigma ^2}}}} \right)$ to denote a
Gaussian PDF with mean $\mu$ and variance $\sigma^2$, and use
$\mathcal{U}({x_i};\frac{1}{N})$ to denote a discrete uniform  PDF
with $N$ points. Finally,  we use notation ${\left\langle \underline
v \right\rangle} \equiv \frac{1}{N}\sum\nolimits_{i = 1}^N {v_i}$
for the sample mean of a certain vector $\underline v \in
\mathbb{R}^N$ and $\eta'( \cdot ) \equiv \frac{\partial }{{\partial
\rho }}\eta(\cdot)$ for the first derivative of the function $\eta(
\cdot )$.

\section{AMP Algorithms:\\Fundamental and Related Works}
The AMP algorithm was originally proposed by Donoho \emph{et al.} to
solve the CS recovery problem with direct sparsity
\cite{AMP1},\cite{AMP2}. The AMP solution $\widehat{\underline
X}^{(t)}=\underline \mu ^{(t)}$ is found by a simple iteration
according to: for the iteration index $t =0,1,2,...$,
\begin{align}\label{standardAMP}
&{\underline \mu  ^{(t + 1)}} = \eta ({{\mathbf{H}}^T}{\underline r ^{(t)}} + {\underline \mu  ^{(t)}})  \\
&{\underline r ^{(t)}} = \underline y  - {\mathbf{H}}{\underline \mu
^{(t)}} + {\underline r ^{(t -
1)}}\begin{array}{l}\frac{N}{M}\end{array}\left\langle { \eta'
({{\mathbf{H}}^T}{{\underline r }^{(t - 1)}} + {{\underline \mu
}^{(t - 1)}})} \right\rangle \nonumber
\end{align}
where   ${\underline r^{(t)}} \in \mathbb{R}^M$ denotes a residual
vector  measuring  fidelity from  $\underline y$ at hand, and
$\eta(\cdot) :\mathbb{R}^N \to \mathbb{R}^N$ indicates a denoising
function, simply called \emph{denoiser}, to realize the sparse
signal estimate $\underline \mu  ^{(t)}\in \mathbb{R}^N$. It is
known that the AMP iteration \eqref{standardAMP} achieves the PT
performance equivalent to that of the Lasso method in the large
limit of $N,M \to \infty$ and $t \to \infty$
\cite{AMP1},\cite{Montanari},\cite{Bayati}. In addition, the AMP
algorithm is basically low-computational with $\mathcal{O}(MN)$
per-iteration cost. Motivated by such excellent properties,
recently, there have been several AMP extensions with the various
types of sparsity:
\begin{itemize}
\item For FD sparsity:  TVAMP \cite{TV_AMP}, AMP with non-local mean  denoiser \cite{NLMAMP}, ssAMP-BGFD \cite{ssAMP1},
\item For group sparsity: Block-AMP \cite{TV_AMP},\cite{BAMP}
\item For complex valued sparsity: Complex-AMP \cite{CAMP}
\item For wavelet sparsity: AMP with amplitude-scale-invariant Bayes'estimator \cite{AMP-ABE}, Turbo-AMP with hidden Markov tree \cite{AMP-wavelet}
\item For generalized sparsity: GrAMPA \cite{GrAMPA}
\end{itemize}

The practical use of the AMP algorithms is not straightforward. This
is mainly caused by the fact that AMP basically postulates  the
matrix entries $h_{ji}\in\mathbf{H}$ following the \emph{i.i.d.}
statistics of $\mathbb{E}[h_{ji}]=0$ and
$\text{Var}[h_{ji}]=\frac{1}{M}$. This postulation is essential to
guarantee the  AMP convergence and validate the state evolution
method providing the own MSE prediction
\cite{Montanari},\cite{Bayati},\cite{Bayati2}. However, the
postulation largely limits practical applicability of the AMP
algorithms for the three main reasons as given below:
\begin{itemize}
\item {\bf{Statistical inconsistency in small systems}}: The law of large number does not hold with small $(M,N)$ such that the sample mean and variance of $h_{ji}$
may not be consistent with $\mathbb{E}[h_{ji}]=0$ and
$\text{Var}[h_{ji}]=\frac{1}{M}$.
\item {\bf{No fast implementation of matrix-vector multiplication}}: The matrix-vector multiplication takes $\mathcal{O}(MN)$ complexity, which  is a  computational  bottleneck in AMP.
This can be relaxed by fast implementation methods, such as
\emph{Fast Fourier Transform} (FFT), when $\mathbf{H}$ is some
unitary  or Toeplitz matrices. In such cases, the complexity is
reduced to $\mathcal{O}(N \log N)$ \cite{Bajwa}. However, there are
no such methods for the \emph{i.i.d.}-random matrices.
\item {\bf{Large memory for storage}}: In general, the pure \emph{i.i.d.}-random matrices $\mathbf{H}$
densely include $\mathcal{O}(MN)$ independent random variables.
Hence, its matrix storage requires significant space with large
$(M,N)$.
\end{itemize}

To overcome these limitations, sub-sampled unitary matrices, such as
\emph{Discrete Cosine Transform} (DCT) matrix, have been tested with
AMP, reporting successful results for implementation and performance
both \cite{EM-BG-GAMP},\cite{AMP_VLSI}. As another direction, some
researchers have attempted to operate the AMP algorithms with
generic matrices $\mathbf{H}$ by using ``damping"
\cite{Vila},\cite{Krz1}, ``mean-removing" \cite{Vila}, ``serial
updating" \cite{Krz2}, and ``free-energy minimization"
\cite{Rangan2}. These approaches generally improve stability of the
AMP convergence at the expense of its recovery speed.

In \cite{GAMP}, Rangan extended the standard AMP \eqref{standardAMP}
to signals $\underline X$, whose elements are  drawn from
generalized  \emph{i.i.d.} PDFs, by applying Bayesian philosophy,
solidifying the foundation for Bayesian AMP works:
\cite{EM-BG-GAMP}-\cite{GrAMPA},\cite{ssAMP1},\cite{Vila},\cite{Rangan2},\cite{AMP-wavelet}.
In the work of \cite{GAMP}, Rangan classifies the Bayesian AMP into
two modes according to its signal estimation criterion.
\begin{itemize}
\item Max-sum mode: The mode is originated from the max-sum loopy BP for the MAP estimation of $\underline X$.
Therefore, the denoiser for this mode estimates the signal by solving a sub-optimization  defined as
\begin{align}\label{denoiser1}
\begin{array}{l}
\eta (\underline \rho  ) \equiv \arg \mathop {\min
}\limits_{\underline X } \frac{1}{2}||\underline \rho   - \underline
X ||_2^2 + g(\mathcal{T}\underline X ).
\end{array}
\end{align}
\item Sum-product mode: The mode is based on the sum-product loopy BP for the MMSE estimation of $\underline X$.
Hence, the denoiser for this mode generates the signal by solving a sub-optimization given as
\begin{align}\label{denoiser2}
\eta (\underline \rho  ) \equiv \frac{1}{Z}\int {\underline x \exp
\left(-\frac{1}{2}||\underline \rho  - \underline x ||_2^2 -
g(\mathcal{T}\underline x )\right)d\underline x }.
\end{align}
\end{itemize}
In \eqref{denoiser1} and \eqref{denoiser2}, $\underline \rho \in
\mathbb{R}^N$ is the denoiser input,   $\mathcal{T}$ is an analysis
operator, and $Z>0$ is a normalization constant. In the Bayesian
AMP,  the regularizer is a functional of the signal prior
$f_{\underline X}(\mathcal{T}\underline x)$, \emph{i.e.}, $g:V \to
\mathbb{R}$ where $V \equiv \{ {f_{\underline X }}:{\mathbb{R}^N}
\to [0,1]\}$, which controls the denoising behavior for enhancing
the signal sparsity in the domain of $\mathcal{T}$.

In the remaining of this section, we briefly introduce the two
existing AMP algorithms applicable to the CS recovery with 1D-FD
sparsity: TVAMP \cite{TV_AMP} and GrAMPA \cite{GrAMPA}, by focusing
on their denoisers $\eta(\cdot)$. These two algorithms will be
included for the simulation-based comparison to the proposed AMP
algorithm, ssAMP-BGFD, in Section IV.

\subsection{TVAMP Algorithm}
Donoho \emph{et al.} introduced TVAMP  for the present problem.
\cite{TV_AMP}.  TVAMP uses the standard AMP iteration, given in
\eqref{standardAMP}, with an anisotropic TV denoiser. TVAMP is
classified to the ``max-sum" mode such that its denoiser can be
represented in the form of the MAP estimation \eqref{denoiser1},
\emph{i.e.},
\begin{align}\label{etaTV}
\begin{array}{l}
\eta (\underline \rho ) \equiv \arg \mathop {\min
}\limits_{\underline X } \frac{1}{2}\left\| {\underline \rho -
\underline X } \right\|_2^2 + \lambda ||\mathbf{D}\underline X||_1,
\end{array}
\end{align}
where the sparsity is enhanced with the $l_1$-regularizer such that
$g(\mathcal{T}\underline X) \equiv \lambda||\mathbf{D}\underline
X||_1$, meaning  from a Bayesian viewpoint that Laplacian prior is
imposed. The implementation denoiser of \eqref{etaTV} requires an
numerical TV minimizer, such as FLSA \cite{FLSA}, TVDIP
\cite{TVDIP}, FISTA-TV \cite{FISTA}, and Condat's direct method
\cite{condat}, since the optimization \eqref{etaTV} is neither
scalar-separable nor smooth such that no closed-form solutions
exist. Therefore, complexity of TVAMP highly depends upon that of
the numerical  minimizer.

\begin{figure}[!t]
\centering
\includegraphics[width=6cm]{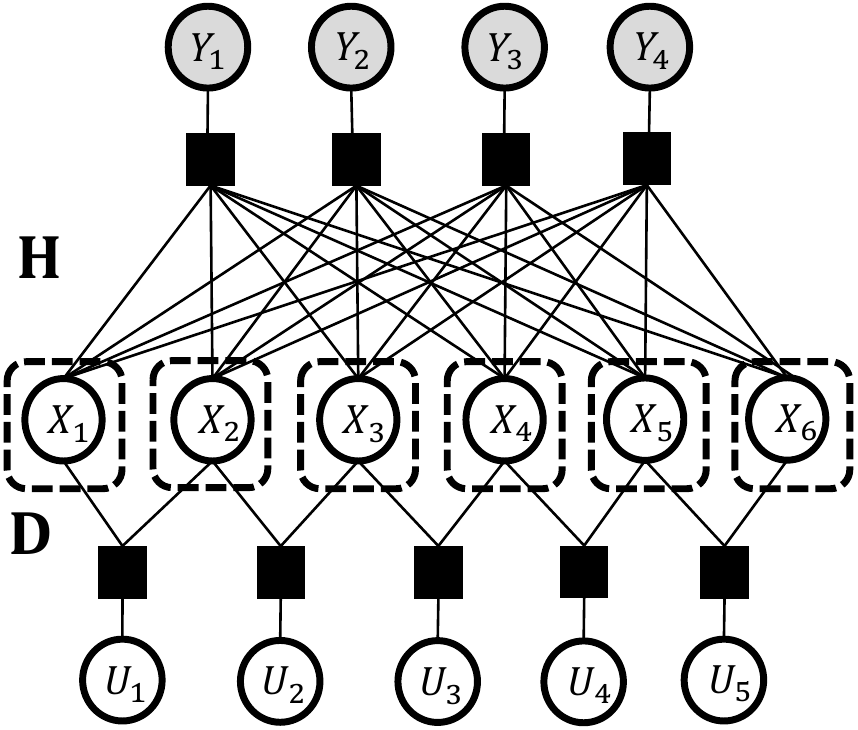}
\caption{Factor graphical modeling of the linear system
\eqref{system}  having a PWC solution $\underline X$, used by the
two AMP algorithms: ssAMP-BGFD (proposed) and the GrAMPA
\cite{GrAMPA}, where the denoiser $\eta(\cdot)$, indicated by
dotted-line boxes in this figure, scalar-separablely works.}
\label{graph_ssAMP}
\bigskip
\centering
\includegraphics[width=6cm]{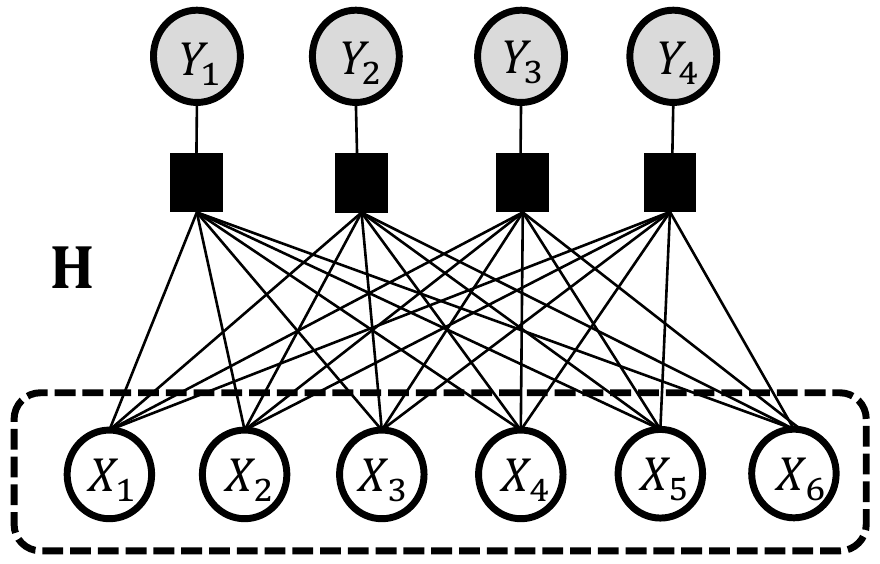}
\caption{Factor graphical modeling of the linear system
\eqref{system} having a PWC solution $\underline X$, used by the
TVAMP algorithm \cite{TV_AMP}, where the non-scalar-separable
denoiser $\eta(\cdot)$, indicated by a dotted-line box in this
figure, is a batch vector function. } \label{graph_TVAMP}
\end{figure}

The  minimizer for \eqref{etaTV} requires batch vector computation,
leading to the non-separability of the TV denoiser. Namely, the
denoising operations is not coordinatewise  as illustrated in
Fig.\ref{graph_TVAMP}. This non-separability prevents TVAMP from
characterizing its MSE in terms of a scalar equivalent model, which
deprives TVAMP of mathematical completeness for its state evolution
formalism \cite{Montanari},\cite{Bayati},\cite{TV_AMP}. The
non-separability does not mean that TVAMP is not scalable for
large-scale problems. TVAMP can have very good scalability for large
$N,M$ according to choice of the numerical TV minimizer (see Section
IV-C for its validation).

\subsection{GrAMPA Algorithm}
Recently, Schniter \emph{et al.} introduced the GrAMPA algorithm for
the analysis CS  setup \cite{GrAMPA}. GrAMPA is useful for general
CS recovery problems with arbitrary analysis operators
$\mathcal{T}$, arbitrary independent signal priors
$g(\mathcal{T}\underline x )\propto \sum\nolimits_d
{g_d([\mathcal{T}\underline x]_d )}$, and arbitrary independent
likelihood $\sum\nolimits_j {\log f_{Y_j}(y_j|{{[\mathbf{H}
\underline x]}_j})}$. Namely, GrAMPA has very good universality to
signal and noise models. Hence, this GrAMPA framework can be simply
configured for the present problem by setting
$\mathcal{T}=\mathbf{D}$ and by assuming the AWGN model. The
corresponding factor graph model is shown in Fig.\ref{graph_ssAMP}.

GrAMPA supports the both modes of the Bayesian AMP. In the present
work, we are interested in the ``sum-product" mode for the method
\eqref{BMMSE}, therefore focusing on the MMSE-based denoiser
 in the form of \eqref{denoiser2}, expressed as
\begin{align}\label{grampa_denoiser1}
\eta (\underline \rho  ) \equiv \frac{1}{Z}\int {\underline x \exp
\left(  -\frac{1}{2}||\underline \rho  - \underline x ||_2^2 -
 \sum\limits_{d = 1}^{N - 1} g_d([\mathbf{D}\underline
x]_d;\widehat{U}_d) \right) d\underline x },
\end{align}
where the authors suggests to use the regularizer
$g_d([\mathbf{D}\underline x]_d;\widehat{U}_d)$ with an MMSE
estimate of a FD scalar:
\begin{align}\label{grampa_denoiser2}
\widehat{U}_d \equiv{\mathbb{E}}\left[ U_d |
[\mathbf{D}\widehat{\underline x}]_d,v_d \right]=\frac{
[\mathbf{D}\widehat{\underline x}]_d }{1 + \omega
\mathcal{N}(0;[\mathbf{D}\widehat{\underline x}]_d,{\nu _d})}.
\end{align}
In \eqref{grampa_denoiser1} and \eqref{grampa_denoiser2}, we define
the random variable $U_d \in \mathbb{R}$ as a clean FD scalar,
assuming that the current signal estimate $\widehat{\underline X}$
is noisy such that $[\mathbf{D}\widehat{\underline X}]_d=U_d +W'$
where $W' \sim \mathcal{N}(0,\nu_d)$. The authors named this
denoiser with \eqref{grampa_denoiser1} and \eqref{grampa_denoiser2}
as the \emph{SNIPE denoiser}. The SNIPE denoiser can support any
Bernoulli-* prior, where ``*" is any continuous PDF, by controlling
the parameter $\omega > 0$.

The naive per-iteration cost of GrAMPA is
$\mathcal{O}(N^2+MN-N)\approx \mathcal{O}(N^2)$ because GrAMPA
operates by the GAMP package with an augmented linear transform
$\mathbf{H}'=\left[ {\frac{{\mathbf{H}}}{{\mathbf{D}}}} \right] \in
\mathbb{R}^{M+N-1 \times N}$. However, its complexity is simply
reduced to  $\mathcal{O}(MN)$ using a fast sparse multiplication
method to $\mathbf{H}'$ \cite{sparse}.

\section{Proposed Algorithm}
In this section, we introduce the proposed algorithm, ssAMP-BGFD,
for solving \eqref{BMMSE} We describe the details of the algorithm
construction: from its factor graphical modeling  to the AMP
approximation. Then, we finalize this section with  discussion about
the prior parameter learning by an EM-tuning method. The overall
 iteration of ssAMP-BGFD is summarized in Algorithm \ref{algo2}.

\subsection{Factor Graphical Model and Prior Model}
The statistical dependency of linear systems  have been effectively
modeled using factor graphs \cite{factor}. Let
$\mathcal{V}\equiv\{1,...,N\}$ be a variable set whose element $i
\in \mathcal{V}$ corresponds to a signal scalar $X_i$, and
$\mathcal{F}_m\equiv\{1,...,M\}$ be a factor set whose element $j
\in \mathcal{F}_m$ corresponds to a measurement scalar $Y_j$. To the
problem, we include another factor set, defined as
$\mathcal{F}_s\equiv\{1,...,N-1\}$, to describe statistical
connections of a finite-difference (FD) scalar $\forall d \in
\mathcal{F}_s: U_d=[\mathbf{D}\underline X]_d$. In order to clarify
two different factors, we name the set $ \mathcal{F}_m$ as
\emph{m-factor} set, and  the set $\mathcal{F}_s$ as \emph{s-factor}
set. Then,  a factor graph, denoted by
$\mathcal{G}(\mathcal{V},{\mathcal{F}_m},{\mathcal{F}_s})$, fully
models the linear system \eqref{system} with a 1D-PWC solution
$\underline X$, as shown in Fig.\ref{graph_ssAMP}. This graph
modeling approach enables us to devise a message-passing rule for
statistically connected signals, which is related to  the approach
of Hybrid-GAMP \cite{HGAMP} and also used in GrAMPA \cite{GrAMPA}.
In addition, for convenience, we indicate the neighboring relation
between the two sets, $\mathcal{V}$ and $\mathcal{F}_s$, by defining
$\forall i \in\mathcal{V}: ne(i)\equiv\{d_1,d_2\in
\mathcal{F}_s|d_1=i-1,d_2=i\}$ and $ \forall d \in\mathcal{F}_s:
ne(d)\equiv\{i_1,i_2\in \mathcal{V}|i_1=d,i_2=d+1\}$.


Based on the graph model designed above, we represent the joint PDF
of the linear system \eqref{system}  as
\begin{align}\label{jointPDF}
{f_{\underline X ,\underline Y }}(\mathbf{D}\underline x ,\underline
y ) = \frac{1}{Z}\underbrace{\prod\limits_{d \in
\mathcal{F}_s}{{f_{U_d}}(u_d ) }}_{\text{Prior}} \underbrace{ \prod
\limits_{j \in \mathcal{F}_m}{{f_{{Y_j}|\mathbf{H}\underline X
}}({y_j}|\mathbf{H}\underline x)}}_{\text{Likelihood}}
\end{align}
where $Z\in \mathbb{R}$ is a normalization constant to validate
$\int {{f_{\underline X ,\underline Y }}(\mathbf{D}\underline x
,\underline y )d\underline x d\underline y  = 1} $. To each m-factor
$j\in \mathcal{F}_m$, we consider an independent Gaussian likelihood
function, \emph{i.e.},
\begin{align}\label{likelihood}
{{f_{{Y_j}|\underline X }}({y_j}|\mathbf{H}\underline x
)}=\mathcal{N}(y_j;\left(\mathbf{H}\underline x\right)_j,\Delta),
\end{align}
for our AWGN noise model where $\Delta$ is the noise variance.

To each s-factor $d\in \mathcal{F}_s$, we impose an independent
Bernoulli-Gaussian (BG) prior for a FD scalar, $\forall i_1,i_2 \in
ne(d): u_d=[\mathbf{D}\underline x]_d=x_{i_1}-x_{i_2}$, by assuming
its sparsity. The BG prior takes a form of the spike-and-slab PDFs
\cite{spikeandslab}, which is given as
\begin{align}\label{prior}
{f_{U_d}}(u_d=x_{i_1}-x_{i_2}) = (1 - q)\delta(u_d) +
q\mathcal{N}(u_d;0,\sigma_0^2 )
\end{align}
where $\delta(u_d)$ denote a Dirac function peaked at $u_d=0$, $q
\in [0,1]$ is a probability weight, and $\sigma_0^2 \in (0,\infty) $
is the variance of the Gaussian PDF.
Following that, the number of nonzeros in FD of $\underline X$ is
explicitly Binomial random with $K\sim\mathcal{B}(N-1,q)$. Such a BG
prior \eqref{prior} has been widely used  in the CS literature with
respect to Bayesian algorithms
\cite{EM-BG-GAMP}-\cite{GrAMPA},\cite{TSW-CS}-\cite{CS-BP2} because
\begin{enumerate}
\item the PDF has a sparsifying ability,
\item the integration of the PDF is  tractable with its
Gaussianity, and
\item the PDF is simply parameterized.
\end{enumerate}
Although one recent paper \cite{antiBG} pointed out that the BG
prior PDF is not appropriate for dealing with discretized
continuous-time signals due to its fast decayed tail, we argue that
the BG prior is still a powerful choice for parametric algorithms,
which keep track a set of statistical parameters such as mean and
variance, by its analytical tractability
\cite{EM-BG-GAMP},\cite{EM-BP}.

\begin{algorithm}[!t]
\caption{Sum-Product Rule}\label{algo1}
\begin{algorithmic}[0]
 \small
\For{$t=1$ \textbf{to} $t^*$}

\State {\bf{ Variable to s-factor (V2sF) update}}

\State $\forall (d,i) \in \mathcal{F}_s \times \mathcal{V} \text{
and } d,d' \in ne(i):$

\State $\widehat v_{i \to d}^{(t + 1)}({x_i})=
\frac{1}{\widetilde Z_{i \to d}}
 s_{d' \to i}^{(t)}({x_i}) \prod\limits_{j\in \mathcal{F}_m} {m_{j \to
i}^{(t)}({x_i})}\,\, (d \ne d')$

\\
\State {\bf{  s-factor to variable (sF2V) update}}

\State $\forall (d,i) \in \mathcal{F}_s \times \mathcal{V} \text{
and } i,i' \in ne(d):$

\State $s_{d \to i}^{(t)}({x_i})= \mathbb{E}_{\widehat v_{i' \to d}^{(t)}(x_i') } \left[{f_{U_d}}(u_d={x_i}-{x_{i'}})|x_{i'}\right]\,\,(i' \ne i )$

\\
\State {\bf{ Variable to m-factor (V2mF) update}}

\State $\forall (j,i) \in \mathcal{F}_m \times \mathcal{V}:$

\State $v_{i \to j}^{(t + 1)}({x_i}) =  \frac{1}{Z_{i \to j}}
\prod\limits_{d \in ne(i)} {s_{d \to i}^{(t)}({x_i})}
\prod\limits_{j' \ne j} {m_{j' \to i}^{(t)}({x_i})}$

\\
 \State {\bf{  m-factor to
variable (mF2V) update}}

\State $\forall (j,i) \in \mathcal{F}_m \times \mathcal{V}:$

\State $m_{j \to i}^{(t)}({x_i}) = \mathbb{E}_{ \{ v_{i' \to j}^{(t)}(x_i') \}} \left[{f_{{Y_j}|\mathbf{H}\underline X }}({y_j}|\mathbf{H}\underline x)| {\{x_{i'}\}} \right]\,\,(i' \ne i )$

\EndFor
\end{algorithmic}
\end{algorithm}

\subsection{Sum-Product Belief Propagation for MMSE Estimation}
We now make use of the factor graphical model of
Fig.\ref{graph_ssAMP} to derive an efficient recovery algorithms for
the present problem. We approach the problem through the MMSE
method, which lead us to the ``sum-product" rule of loopy BP
\cite{GAMP},\cite{Bishop}. There exists a vast literature justifying
the use of the sum-product BP algorithm, applying them on concrete
problems such as channel coding \cite{Gallager}, computer vision
\cite{non_para_BP}, as well as compressed sensing (CS)
\cite{CS-BP2},\cite{SuPrEM}.

\begin{figure}[!t]
\centering
\includegraphics[width=8.5cm]{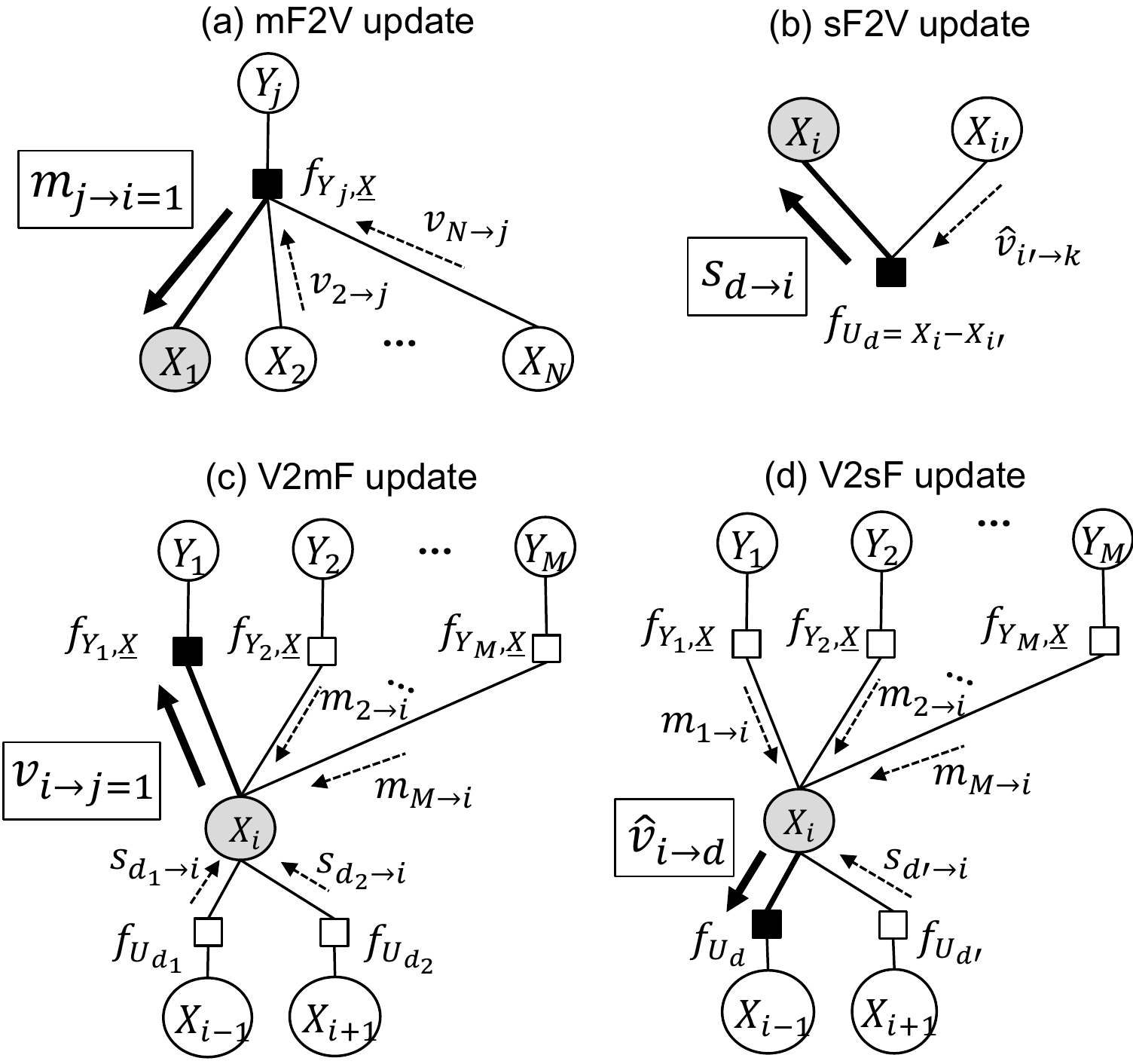}
\caption{Factor graphical representation of the sum-product rule:
(a) m-factor to variable (mF2V) update, (b) s-factor to variable
(sF2V) update, (c) variable to m-factor (V2mF) update, (d) variable
to s-factor (V2sF) update.} \label{fig1-2} \vspace{-10pt}
\end{figure}

We construct a sum-product rule based on the joint PDF of
\eqref{jointPDF}, which consists of four types of the local message
updates as graphically illustrated in Fig.\ref{fig1-2} and listed in
Algorithm \ref{algo1}, where the expectation of the sF2V and mF2V
updates are over the previous V2sF and V2mF messages, respectively;
the constants $Z_{ i \to j}, \widetilde Z_{i \to d} >0$ are for
normalization. This sum-product task is divided into two parts:
\begin{enumerate}
\item Pursuing the 1D-FD sparsity with the independent BG prior \eqref{prior},
\item Seeking
the measurement fidelity with the independent Gaussian likelihood
function \eqref{likelihood} for the AWGN model.
\end{enumerate}
The first part  is with respect to the s-factors $\mathcal{F}_s$
(the V2sF and sF2V updates), and the second part is with the
m-factors $\mathcal{F}_m$ (the V2mF and mF2V updates). Then, at the
fixed-point ($t=t^*$), the marginal posterior of $X_i$ is
approximated by
\begin{align}\label{marginalPDF}
& f_{X_i|\underline Y}(x_i|\underline y) \propto
\prod\limits_{d \in ne(i)} {s_{d \to i}^{(t=t^*)}({x_i})}
\prod\limits_{j\in\mathcal{F}_m} {m_{j \to i}^{(t=t^*)}({x_i})}.
\end{align}
Using \eqref{marginalPDF}, we provide an MMSE approximation of
$\widehat{X_i}$, whose function is defined as the denoiser of the
ssAMP-BGFD algorithm, \emph{i.e.},
\begin{align}\label{marginal_mean}
\mu_i={\eta}( \cdot ) &\equiv \mathbb{E}_{f_{X_i|\underline Y}}\left[{X_i}|{\mathbf{H}},\underline Y =\underline y \right],
\end{align}
and the corresponding variance function is given as
\begin{align}\label{marginal_var}
\sigma_i^2={\gamma}( \cdot ) &\equiv \mathbf{Var}_{f_{X_i|\underline Y}}\left[{X_i}|{\mathbf{H}},\underline Y =\underline y \right].
\end{align}
However, as claimed in literature \cite{AMP1}-\cite{EM-BP},
Algorithm \ref{algo1} is infeasible in practice because \emph{i)}
the messages are density function over the real line, and \emph{ii)}
$2MN+4(N-1)$ message exchanges are required per iteration.

\subsection{AMP Approximation}
We approach  the computational infeasibility of Algorithm
\ref{algo1} via the AMP approximation, which have been discussed and
analyzed in the literature \cite{AMP1}-\cite{EM-BP},\cite{Bayati}.
The AMP approximation produces a remarkably simpler algorithm, whose
messages are real numbers instead of density functions, which
handles $\mathcal{O}(M+N)$ messages rather than $\mathcal{O}(MN)$
messages per iteration. In the conventional literature
\cite{AMP1}-\cite{EM-BP}, this approximation consists of two steps:
\begin{itemize}
\item {\bf{Parameterization step}}:  Based on the central limit theorem (CLT), the sum-product rule is approximated to a
parametric message-passing rule exchanging a pair of real
numbers,
\item {\bf{First-order
approximation step}}: This step cancels interference caused by the
loopy graph connection, leading to reduction of  the number of the
messages handled in the mF2V and V2mF updates.
\end{itemize}
In addition to these steps, the present work includes the third
step, called \emph{Right/Left Toward Message-Passing} (R2P/L2P)
step, to deal with the sF2V and V2sF updates. This R2P/L2P update is
devoted to promote the signal sparsity over the statistical chain
connection with $\mathcal{V}$ and $\mathcal{F}_s$.

Throughout this AMP approximation, we assume that the matrix
$\mathbf{H}$ is a dense \emph{i.i.d.}-random matrix, \emph{i.e.},
its entries $h_{ji}\in \mathbf{H}$ are randomly distributed with
zero mean and variance $\frac{1}{M}$; hence,
${\mathbb{E}}||\underline h_i ||_2^2 = 1$. In addition, we clarify
beforehand that this AMP approximation is heuristic.  Namely, we do
not claim any mathematical equivalence between the sum-product rule
of Algorithm \ref{algo1} and the ssAMP-BGFD rule produced by this
approximation.

\subsubsection{STEP I - Parameterization Step} We begin this
step with definitions of the mean and variance of $X_i$ over the
message densities:
\begin{subequations}\label{mean_and_var}
\begin{eqnarray}
\mu _{i \to j} &=&\mathbb{E}_{v_{i \to j}}[X_i|{\mathbf{H}},\underline Y =\underline y ],\label{V2mfmean}\\
\sigma _{i \to j}^2 &=& \mathbf{Var}_{v_{i \to j}}[X_i|{\mathbf{H}},\underline Y =\underline y ],\label{V2mfvar}\\
\mu _{i \to d} &=& \mathbb{E}_{\widehat v_{i \to
d}}[X_i|{\mathbf{H}},\underline Y =\underline y ],\\
\sigma _{i \to d}^2 &=& \mathbf{Var}_{\widehat v_{i \to
d}}[X_i|{\mathbf{H}},\underline Y =\underline y ].
\end{eqnarray}
\end{subequations}
For large $N$, we can approximate exponent of the mF2V message by a
quadratic function based on CLT; then, the mF2V message becomes a
scaled Gaussian PDF \cite{GAMP},\cite{EM-BP}. The sF2V message is
represented as a Bernoulli-Gaussian PDF by calculating the
integration with the BG prior \eqref{prior}. Using these two facts,
we specify the message representation from Algorithm \ref{algo1}:
\begin{itemize}
\item V2sF messages:
\begin{align}\label{pp_v2sf}
\begin{array}{l}
{\widehat v_{i \to d}}({x_i}) =\frac{1}{\widetilde Z_{i \to d}}{s_{d' \to
i}}({x_i})\mathcal{N}({x_i};{\rho _{i}},\theta_i ),
\end{array}
\end{align}
\item  sF2V  messages:
\begin{align}\label{pp_sf2v}
\begin{array}{l}
{s_{d \to i}}({x_i}) =(1 - q) \mathcal{N} ({x_i};{\mu
_{i' \to d}},{\sigma _{i' \to d}^2})\\
\,\,\,\,\,\,\,\,\,\,\,\,\,\,\,\,\,\,\,\,\,\,\,\,\,\,\,\,\,\,\,\,\,\,\,\,\,\,\,\,+ q\mathcal{N} ({x_i};{\mu _{i'
\to d}},\sigma _0^2 + \sigma _{i' \to d}^2 ),
\end{array}
\end{align}
\item V2mF  messages:
\begin{align}\label{pp_v2mf}
\begin{array}{l}
{v_{i \to j}}({x_i}) = \frac{1}{Z_{i \to j}}\prod\limits_{d \in ne(i)} {s_{d \to i}({x_i})}
{\mathcal{N}({x_i};{\rho _{i\to j}},\theta_{i\to j} )},
\end{array}
\end{align}
\item mF2V messages:
\begin{align}\label{pp_mf2v}
\begin{array}{l}
{m_{j \to
i}}({x_i}) \propto \mathcal{N}\left(h_{ji}{x_i};r_{j \to i}, \theta_{j \to i} \right),
\end{array}
\end{align}
\end{itemize}
where we need  several parameter definitions:
\begin{subequations}\label{para1}
\begin{eqnarray}
{\rho _{i \to j}} &\equiv& \sum\nolimits_{j' \ne j} {h_{j'i}}{r_{j' \to i}},\label{rho_i} \\
{\rho _{i }} &\equiv&  \sum\nolimits_{j \in \mathcal{F}_m} {h_{ji}}{r_{j \to i}},\label{rho_i2}   \\
{r_{j \to i}} &\equiv& {y_j} - \sum\nolimits_{i' \ne i} {h_{ji'}^{}\mu _{i' \to j}^{}},\\
{\theta _{j \to i}} &\equiv& \Delta  + \sum\nolimits_{i' \ne i} h_{ji'}^2 \sigma _{i' \to j}^2, \label{theta_mf2v}
\end{eqnarray}
\end{subequations}
and in the large limit $(M\to \infty)$, the variance parameter ${\theta _{i \to j}}$ can drop its directional nature, \emph{i.e.},
\begin{align}\label{para2}
{\theta _{i \to j}} \equiv \sum\limits_{j' \ne j} {h_{j'i}^2{\theta _{j' \to i}}} \mathop  = \limits^{M \to \infty } \frac{1}{M}\sum\limits_{j\in \mathcal{F}_m} {{\theta _{j \to i}}}  = {\theta _i}.
\end{align}
Equations \eqref{mean_and_var},\eqref{para1},\eqref{para2} establish
a message update rule which only exchanges the parameters of the
message densities \eqref{pp_v2sf}-\eqref{pp_mf2v}: namely,
$(\rho_i,\theta_i,\{\mu _{i \to d},\sigma _{i \to d}^2 \}_{d\in
ne(i)})$ for the sF2V and V2sF updates, $(\rho_{i \to
j},\theta_{i\to j}, \mu_{i \to j},\sigma^2_{i \to j})$ for the V2mF
update, and $(r_{ j \to i}, \theta_{j \to i})$ for the mF2V update.

To formulate the calculations of \eqref{mean_and_var}  with the parameters we
have defined in \eqref{para1},\eqref{para2},  we further define
\begin{subequations}\label{func}
\begin{align}
&\eta(\cdot)\equiv\mathbb{E}_{v_{i \to j}}[X_i|\rho_{i \to j}, \theta_i, \{\mu _{i \to d},\sigma _{i \to d}^2 \}_{d\in ne(i)}],\label{func1}\\
&\gamma(\cdot)\equiv\mathbf{Var}_{v_{i \to j}}[X_i|\rho_{i \to j}, \theta_i, \{\mu _{i \to d},\sigma _{i \to d}^2 \}_{d\in ne(i)}],\label{func2}\\
&\phi(\cdot) \equiv \mathbb{E}_{\widehat v_{i \to
d}}[X_i|\rho_{i}, \theta_i, \mu _{i \to d'},\sigma _{i \to d'}^2],\\
&\zeta(\cdot)\equiv \mathbf{Var}_{\widehat v_{i \to
d}}[X_i|\rho_{i}, \theta_i, \mu _{i \to d'},\sigma _{i \to d'}^2].
\end{align}
\end{subequations}
The V2mF  calculations of \eqref{V2mfmean},\eqref{V2mfvar} share the
functions, $\eta(\cdot)$ and $\gamma(\cdot)$, with the MMSE
approximation of \eqref{marginal_mean},\eqref{marginal_var},
respectively. This is based on the fact that the marginal posterior
and the V2mF message are equivalent PDFs except the difference of
$\rho_i$ and $\rho_{i \to j}$. We can represent the function
$\eta(\cdot)$ in the form of an MMSE-based denoiser
\eqref{denoiser2}, \emph{i.e.},
\begin{align}\label{ssAMPdenoiser}
&\eta ({\rho _i};{\theta _i},\{\mu _{i' \to d},\sigma _{i' \to d}^2 \}_{d\in ne(i)})\\
&\equiv \frac{1}{Z}\int {{x_i}\exp \left( { - \frac{{{ ({x_i} - {\rho _i})}^2} }{{2{\theta _i}}} - {g_i}({x_i};\{\mu _{i' \to d},\sigma _{i' \to d}^2 \})} \right)d{x_i}} \nonumber
\end{align}
where the FD sparsity regularizer ${g_i}({x_i};\cdot)$ is defined as
\begin{align}
{g_i}({x_i};\{\mu _{i' \to d},\sigma _{i' \to d}^2 \})\equiv  - \sum\limits_{(d,i')} {\log \,{s_{d \to i}}({x_i};\mu _{i' \to d},\sigma _{i' \to d}^2 )}.\nonumber
\end{align}
for $(d,i') \in \{ ({d_1},i - 1),({d_2},i + 1)|d_1,d_2 \in ne(i)\}$.
In addition, we emphasize here that all the functions in
\eqref{func} basically  maps  a scalar input onto  a scalar output.
This property of the functions was introduced that a function is
``scalar-separable" if for a vector input $\underline \rho
=[\rho_1,...,\rho_N]^T \in\mathbb{R}^N$, we have $\eta (\underline
\rho;\cdot ) = {[\eta ({\rho _1};\cdot ),...,\eta ({\rho _N};\cdot
)]^T}\in \mathbb{R}^N$ \cite{TV_AMP}.

The  sF2V message modeling is one main difference of the two AMP
algorithms originated from the same graph
$\mathcal{G}(\mathcal{V},{\mathcal{F}_m},{\mathcal{F}_s})$:
ssAMP-BGFD and GrAMPA. In GrAMPA, the sF2V message is approximated
to a scaled Gaussian PDF as done with the mF2V message. However, the
1D-FD operator $\mathbf{D}$ does not include a sufficient number of
the row weights to hold the law of large numbers for CLT; hence, the
Gaussian approximation of GrAMPA is limited at the s-factor.  In
contrast, ssAMP-BGFD precisely models the sF2V message using a BG
density without any approximation, as shown in \eqref{pp_sf2v}. This
is connected to the faster convergence characteristic of ssAMP-BGFD
(see Section IV-B for empirical validation).

\subsubsection{STEP II - First-Order Approximation at
M-factors} The AMP approximation reduces the number of messages
handled per iteration, by removing directional nature from the V2mF
and mF2V messages. Then, the AMP iteration contains only
$\mathcal{O}(M+N)$ messages over the edges $(i,j) \in \mathcal{V}
\times \mathcal{F}_m$ per iteration, which is much smaller than
$\mathcal{O}(M N)$ of the parameter-passing rule.

The  directional nature of the messages
depends on the index of destination nodes. For instance, the
direction of $\{\mu_{ i \to j}\}_{j \in
\mathcal{F}_m}$, sent by a fixed node $ i \in \mathcal{V}$, is determined
only  by $j \in \mathcal{F}_m$  since the terms, excluded from the sum
on \eqref{rho_i}, are changed by $j \in \mathcal{F}_m$. Therefore, it is natural to represent
the V2mF parameters as
\begin{subequations}\label{direc_corr}
\begin{align}
\mu _{i \to j} &= \mu_i + \Delta \mu_{i \to j},\\
\sigma^2 _{i \to j} &= \sigma^2_i + \Delta \sigma^2_{i \to j},\\
\rho _{i \to j} &=\rho_i +\Delta \rho _{i \to j},\\
\theta _{i \to j} &= \theta _i+\Delta \theta _{i \to j}
\end{align}
\end{subequations}
where  $\Delta \mu_{i \to
j},\Delta \sigma^2_{i \to
j},\Delta \rho_{i \to j},\Delta \theta_{i \to j} \in \mathbb{R}$ are
the directional correction terms having order
$\mathcal{O}(N^{-1/2})$.
We can apply the expressions \eqref{direc_corr} to establish a non-directional V2mF and mF2V  updates, which will lead to
the message reduction.

\begin{figure}[!t]
\centering
\includegraphics[width=8.9cm]{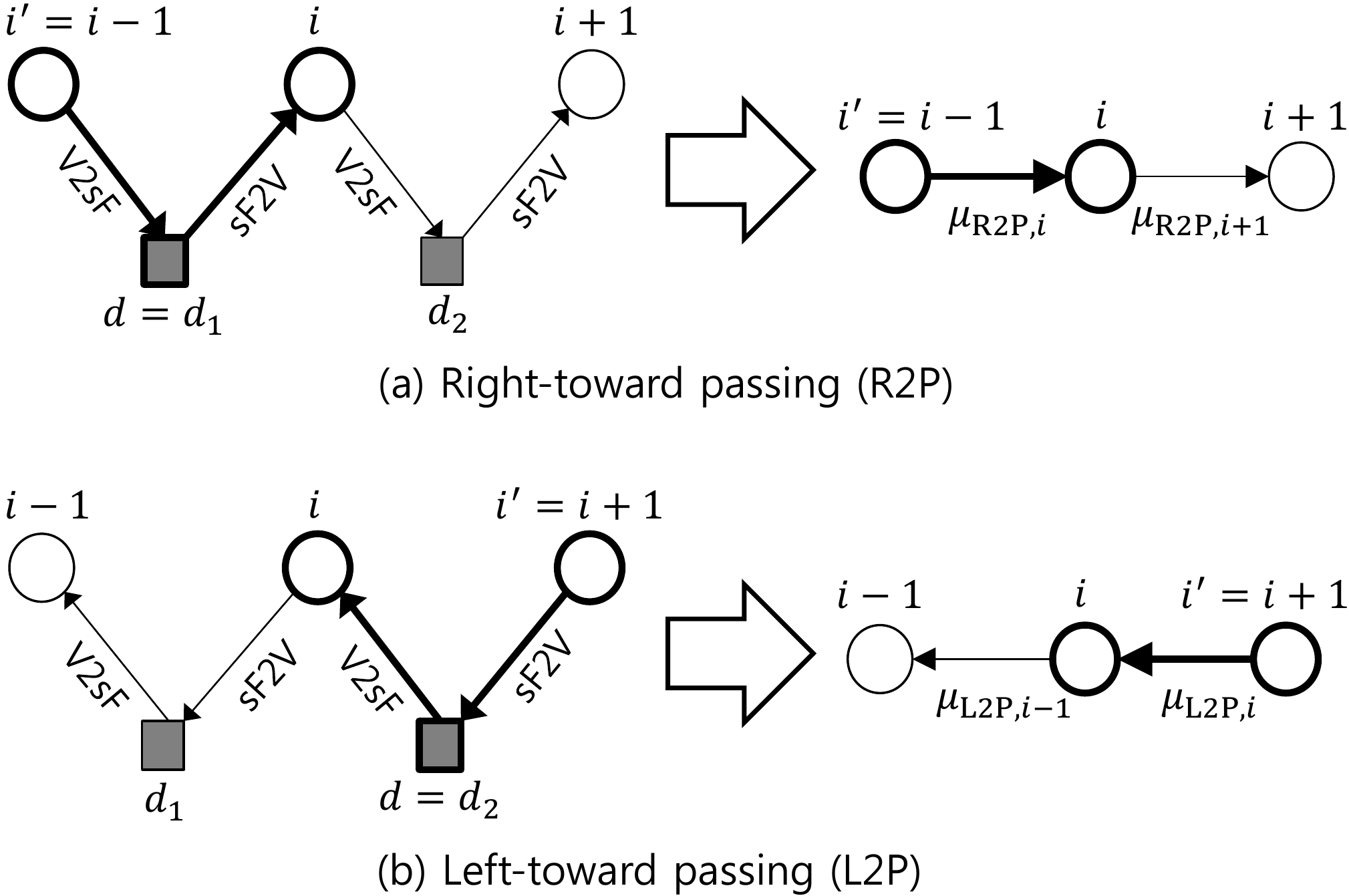}
\caption{Graphical representation of the R2P/L2P update. This
R2P/L2P update enables  ssAMP-BGFD to exchange the adjacent
information over the chain connection with $\mathcal{V}$ and
$\mathcal{F}_s$, which is for seeking sparsity in the
finite-difference $U_d=X_i-X_{i'}$ where $i,i'\in \mathcal{V}$ and
$i,i' \in ne(d)$. } \label{fig2}
\end{figure}

The first key of this approach is to represent the residual,
\begin{align}\label{residual}
{r_j} = {y_j} - \sum\nolimits_{i\in \mathcal{V}} {{h_{ji}}({\mu
_{i}+{\Delta\mu _{i \to j}}  } )}\,\,\forall j \in
\mathcal{F}_m,
\end{align}
as a function of the non-directional parameters for the mF2V update.
If done so, this will lead to a non-directional expression of
$\rho_i,\mu_i$ for the V2mF update. Namely,  from \eqref{rho_i2},
we have
\begin{align}\label{rho2}
{\rho _i}& = \sum\nolimits_{j \in \mathcal{F}_m} {{h_{ji}}( \underbrace{{y_j} - \sum\nolimits_{i' \in \mathcal{V}} {{h_{ji'}}{\mu _{i' \to j}}}}_{=r_j}  + {h_{ji}}{\mu _{i \to j}})} \nonumber  \\
   &= \sum\nolimits_{j \in \mathcal{F}_m} {{h_{ji}}{r_j}}  + \underbrace {\sum\nolimits_{j \in \mathcal{F}_m} {h_{ji}^2{\mu _{i \to j}}} }_{ ={\mu _i} (\text{as } M \to \infty ) },
\end{align}
which is an input of \eqref{ssAMPdenoiser} to generate $\mu_i$. The
mF2V variance $\theta_{j \to i}$ becomes needless since by plugging
\eqref{theta_mf2v} in \eqref{para2}, we can obtain $\theta_i$
directly from the V2mF variance $\sigma_i^2$, \emph{i.e.},
\begin{eqnarray}\label{rho_theta_i}
\begin{array}{l}
{\theta _i} = \Delta  + \frac{1}{M}\sum\limits_{i'\in \mathcal{V}} \underbrace {\sum\nolimits_{j \in \mathcal{F}_m} {h_{ji'}^2\sigma _{i' \to j}^2} }_{ = \sigma _{i'}^2 (\text{as } M \to \infty)}  - \underbrace{\sum\limits_{j \in \mathcal{F}_m} {h_{ji}^2\sigma _{i \to j}^2}}_{= \sigma _i^2 (\text{as } M \to \infty)}  \\
\,\,\mathop  = \limits^{N \to \infty } \Delta  +
\frac{1}{M}\sum\nolimits_{i \in \mathcal{V}} {\sigma _i^2} \equiv
\theta
\end{array}
\end{eqnarray}
where we can drop the index $i\in\mathcal{V}$ from ${\theta}_i$ with
$N \to \infty$. Instead of \eqref{rho_theta_i}, we can use an
approximation \cite{Montanari},\cite{EM-BP}
\begin{align}\label{Delta_appx}
\theta \approx \frac{1}{M}|| \underline r ||^2_2.
\end{align}
In this case, the variance estimation of ${\sigma _i^2} $ is also
not necessary.

Then, the remaining is to obtain an non-directional expression of
\eqref{residual}. We approach this through the two-step manipulation
given below:
\begin{enumerate}
\item Applying the first-order  approximation to the V2mF calculation,  $\mu_{ i \to j}=\eta(\rho_i +\Delta \rho
_{i \to j}; \theta _i + \Delta \theta _{i \to j})$,
\item Substituting the result of the first step to
\eqref{residual}.
\end{enumerate}
This  approach  has been introduced in
\cite{AMP1}-\cite{EM-BP},\cite{Bayati}, where the authors verified
that although approximation errors are induced in the manipulation,
the errors are negligible with the large system limit ($N,M \to
\infty$). We omit  the details of the manipulation by referring the
reader to the  conventional literature
\cite{AMP1}-\cite{EM-BP},\cite{Bayati}. As a result, we obtain a
non-directional expression of \eqref{residual}:
\begin{eqnarray}\label{residual3}
\begin{array}{l}
{r_j}\mathop  = \limits^{N,M \to \infty } {y_j} - {\sum\nolimits_i
{h_{ji}}{\mu _i} + \underbrace{{r_j}\frac{N}{M}\left\langle {\eta
'(\rho _i;\cdot)} \right\rangle }_{\text{Onsager term}} }.
\end{array}
\end{eqnarray}
The last term of \eqref{residual3} corresponds to the term
$\sum\nolimits_i {h_{ji}^{}\Delta {\mu _{i \to j}}} $ of
\eqref{residual}, which corrects the dependency on the index
$j\in\mathcal{F}_m$ in the directional parameter $\mu_{i \to j}$.
This correction term has been called \emph{Onsager term} in the
literature \cite{AMP1}-\cite{EM-BP},\cite{Bayati} which is known as
a key for convergence of the AMP iterations.

\begin{algorithm}[!t]
\caption{ssAMP-BGFD }\label{algo2}
\begin{algorithmic}[0]
\small
\State {\bf{Inputs:}} Measurements $\underline y$, a measurement
matrix $\mathbf{H}$, \\ prior parameters $q, \sigma_0$

 \\
\State {\bf{Initialization:}}

\State set $\{{\underline \mu  },\underline \sigma^2  \}^{(t = 0)} =
\{\underline 0, \underline 1 \sigma_0^2\}$, $\underline r^{(t = 0)}
= \underline y$

\State set $\{{\underline \mu_{\text{R2P}}},\underline
\sigma_{\text{R2P}}^2\}^{(t = 0)} = \{\underline 0, \underline 1
\sigma_0^2\}$, $\{{\underline \mu_{\text{L2P}}},\underline
\sigma_{\text{L2P}}^2\}^{(t = 0)} = \{\underline 0, \underline 1
\sigma_0^2\}$

\\
\State {\bf{Iteration:}} \For{$t=1$ \textbf{to} $t^*$}

\State set ${\underline \rho  ^{(t)}} = {{\mathbf{H}}^T}\underline r
_{}^{(t-1)} + \underline \mu  _{}^{(t - 1)}$ \State set $\theta
_{}^{(t)} = \Delta  + \frac{1}{M} \underline{1}^T (\underline
\sigma^2)^{(t)}$
\\

\State  set $\forall i \in
\mathcal{V}\backslash\{1\}:$\\
$\begin{array}{l}
  \left\{ \begin{gathered}
  {\mu _{{\text{R2P}},i}}, \hfill \\
  \sigma _{{\text{R2P}},i}^2 \hfill \\
\end{gathered}  \right\}^{(t)} \mathop  =  \left\{ \begin{gathered}
  \phi ({\rho _{i - 1}^{(t)}};{\theta ^{(t )}  }, \{{\mu _{{\text{R2P}},i - 1} },\sigma _{{\text{R2P}},i - 1}^2\}^{(t-1)}), \hfill \\
  \zeta ({\rho _{i - 1}^{(t)}};{\theta ^{(t )}  }, \{{\mu _{{\text{R2P}},i - 1}},\sigma _{{\text{R2P}},i - 1}^2\}^{(t-1)}) \hfill \\
\end{gathered}  \right\} \hfill \\
\end{array}$
\\

\State set $\forall i \in
\mathcal{V}\backslash\{N\}:$\\
$\begin{array}{l}
  \left\{ \begin{gathered}
  {\mu _{{\text{L2P}},i}}, \hfill \\
  \sigma _{{\text{L2P}},i}^2 \hfill \\
\end{gathered}  \right\}^{(t)}  \mathop  = \left\{ \begin{gathered}
  \phi ({\rho _{i + 1}^{(t)}};{\theta^{(t )}  },\{{\mu _{{\text{L2P}},i + 1}},\sigma _{{\text{L2P}},i + 1}^2\}^{(t-1)}), \hfill \\
  \zeta ({\rho _{i + 1  }^{(t)}};{\theta^{(t )} },\{{\mu _{{\text{L2P}},i + 1}},\sigma _{{\text{L2P}},i + 1}^2\}^{(t-1)}) \hfill \\
\end{gathered}  \right\}
\end{array}$
\\
\\
 \State set
$\underline \mu^{(t)} = \eta {({{\underline
\rho}^{(t)}};\theta^{(t)},{\underline \mu_{\text{R2P}}},\underline
\sigma_{\text{R2P}}^2,{\underline \mu_{\text{L2P}}},\underline
\sigma_{\text{L2P}}^2 )}$
 \State set
$ (\underline {\sigma}^2)^{(t)} = \gamma {({{\underline
\rho}^{(t)}};\theta^{(t)},{\underline \mu_{\text{R2P}}},\underline
\sigma_{\text{R2P}}^2,{\underline \mu_{\text{L2P}}},\underline
\sigma_{\text{L2P}}^2 )}$
\\

\State  $\begin{gathered}
  \text{set }\underline r _{}^{(t)} = \underline y  - {\mathbf{H}}\underline \mu^{(t)}\hfill\\
  \,\,\,\,\,\,\,\,\,\,\,\,\,\,\,\,\,\,\,\,\,\,\,\,\,+ \underline r _{}^{(t - 1)} \frac{N}{M}{\left\langle {\eta '({\underline \rho  }^{(t)};\theta^{(t)},{\underline \mu_{\text{R2P}}},\underline
\sigma_{\text{R2P}}^2,{\underline \mu_{\text{L2P}}},\underline
\sigma_{\text{L2P}}^2  )} \right\rangle} \hfill \\
\end{gathered} $
\EndFor
\\
\State {\bf{Outputs:}} $\widehat { \underline
x}_{\text{ssAMP-1D}}=\underline \mu ^{(t=t^*)}$
\end{algorithmic}
\end{algorithm}

\subsubsection{STEP III - Right/Left Toward Message-Passing  at
S-factors} In our factor graph model
$\mathcal{G}(\mathcal{V},{\mathcal{F}_m},{\mathcal{F}_s})$,  the
edge connections between $\mathcal{V}$ and $\mathcal{F}_s$ are
stronger than the connections between $\mathcal{V}$ and
$\mathcal{F}_m$. This ``weak/strong'' concept is based on two facts:
\begin{itemize}
\item A s-factor $d\in \mathcal{F}_s$ has only two connections to
$\mathcal{V}$; hence, the corresponding two scalars $X_i, X_{i'}$
($i,i'\in ne(d)$) have potentially larger influence on $d\in
\mathcal{F}_s$ than a certain m-factor $j\in \mathcal{F}_m$ which
has the other $N-2$ connections to $\mathcal{V}\backslash \{i,i'\}$,
\item The edge weight to $\mathcal{F}_s$ is relatively larger than the weight to
$\mathcal{F}_m$: specifically, the edge weight to $\mathcal{F}_s$ is
deterministically `1', whereas the weight to $\mathcal{F}_m$ is
imposed by the matrix entry $h_{ji}\in \mathbf{H}$ which is randomly
distributed with zero-mean and the variance $\frac{1}{M}$.
\end{itemize}
For such  strong edges, the approximation, given in the STEP II,
does not hold \cite{HGAMP}. Therefore, the sF2V and V2sF updates
remains in the conventional sum-product form over the chain
connection with $\mathcal{V}$ and $\mathcal{F}_s$.

Nevertheless, there is still room for the algorithm simplification.
We note from \eqref{pp_sf2v} that the sF2V update $(d \to i)$ is
simple assignment of the V2sF parameters $(i' \to d)$ according to
the direction of message-passing, where $i,i'\in ne(d)$. This
direction is decided by placement of the s-factor $d\in ne(i)$.
\begin{itemize}
\item When $d=d_1$ such that the s-factor is placed on the leftside of
the variable node $i\in \mathcal{V}\backslash\{1\}$, we have
$i'=i-1$; hence, the sF2V parameters is toward right  (see
Fig.\ref{fig2}-(a)).
\item When
$d=d_2$ such that the s-factor is on the rightside of the  node
$i\in \mathcal{V}\backslash\{N\}$, we have $i'=i+1$; hence, the sF2V
parameters is left-toward (see Fig.\ref{fig2}-(b)).
\end{itemize}
Accordingly, what we only need is to keep track the V2sF update
 according to the direction of the sF2V
message-passing. We combine these two updates by defining the
\emph{Right/Left Toward Message-Passing} (R2P/L2P) update as
\begin{align}\label{R2P_L2P}
\small
\begin{array}{l}
  {\text{1)  R2P}}\,{\text{update:}} \hfill \\
\forall i \in \mathcal{V}\backslash\{1\} \text{ and } i'=i-1:\\
 \left\{ \begin{gathered}
  {\mu _{{\text{R2P}},i}}, \hfill \\
  \sigma _{{\text{R2P}},i}^2 \hfill \\
\end{gathered}  \right\} \equiv   \left\{ \begin{gathered}
  \mu _{i-1 \to d}, \hfill \\
  \sigma _{i-1 \to d}^2 \hfill \\
\end{gathered}  \right\}  = \left\{ \begin{gathered}
  \phi ({\rho _{i - 1}};{\theta},{\mu _{{\text{R2P}},i - 1}},\sigma _{{\text{R2P}},i - 1}^2), \hfill \\
  \zeta ({\rho _{i - 1}};{\theta },{\mu _{{\text{R2P}},i - 1}},\sigma _{{\text{R2P}},i - 1}^2) \hfill \\
\end{gathered}  \right\}, \hfill \\\\
  {\text{2)  L2P}}\,{\text{update:}} \hfill \\
\forall i \in \mathcal{V}\backslash\{N\} \text{ and } i'=i+1:\\
    \left\{ \begin{gathered}
  {\mu _{{\text{L2P}},i}}, \hfill \\
  \sigma _{{\text{L2P}},i}^2 \hfill \\
\end{gathered}  \right\} \equiv  \left\{ \begin{gathered}
  \mu _{i+1 \to d}, \hfill \\
  \sigma _{i+1 \to d}^2 \hfill \\
\end{gathered}  \right\} = \left\{ \begin{gathered}
  \phi ({\rho _{i + 1}},{\theta };{\mu _{{\text{L2P}},i + 1}},\sigma _{{\text{L2P}},i + 1}^2), \hfill \\
  \zeta ({\rho _{i + 1  }},{\theta };{\mu _{{\text{L2P}},i + 1}},\sigma _{{\text{L2P}},i + 1}^2) \hfill \\
\end{gathered}  \right\},
\end{array}
\end{align}
where without loss of generality, we set $\{{ \mu_{\text{R2P},i=1}},
\sigma_{\text{R2P},i=1}^2\} = \{{\mu_{\text{L2P},i=N}},
\sigma_{\text{L2P},i=N}^2\}=\{0, \sigma_0^2\}$ for $i= 1, N$. These
R2P/L2P updates take a role of exchanging the neighboring
information over the chain connection with $\mathcal{V}$ and
$\mathcal{F}_s$, promoting the FD sparsity of $\underline X$. It is
clarified from \eqref{R2P_L2P} that ssAMP-BGFD expends
$\mathcal{O}(N)$ per-iteration cost  for the R2P/L2P update.

\subsection{EM-Tuning of  Prior Parameters}
We provide an  online-tuning strategy for the prior parameters,
$\tau\equiv\{q,\sigma_0^2\}$ in ssAMP-BGFD.  For this, we setup an
maximum likelihood estimation (MLE), applying a popular technique,
\emph{Expectation-Maximization} (EM), to  the  estimation. This EM
approach goes well with the Bayesian AMP parameter tuning, which has
been demonstrated by Schniter \emph{et al.} \cite{EM-BG-GAMP},
Kamilov \emph{et al.} \cite{Kamilov}, and Krzakala \emph{et al.}
\cite{EM-BP} for the CS recovery problem with direct sparsity.

\begin{figure}
\centering
\includegraphics[width=5cm]{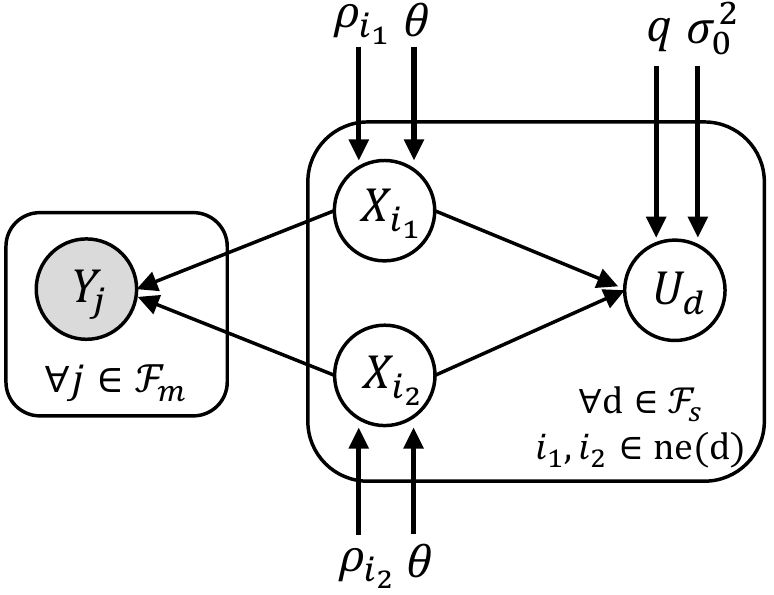}
\caption{Graphical representation of the EM-tuning  of the prior
parameters $\mathbf{\tau}\equiv\{q,\sigma_0^2\}$} \label{fig_EM}
\end{figure}

The statistical dependency in our MLE setup is graphically represented in
Fig.\ref{fig_EM} where  $X_{i_1},
X_{i_2}$ and $\underline{Y}$ are related by the measurement model \eqref{system}, and we know the connection
$i_1,i_2 \in ne(d): U_d=X_{i_2} - X_{i_1}$ from \eqref{prior}.
In this MLE, we consider the evidence PDF $f_{\underline
Y}(\underline y|\tau)$ as the likelihood function. As in \cite{Bishop}, we can decompose the log-likelihood into
\begin{align}\label{EM_LH}
\log f_{\underline Y}(\underline y|{\tau}) =
\mathcal{L}({\widehat f_{{\underline U}}},{\tau}) +
{\text{KL}}(\widehat f_{\underline U}||f_{{\underline U}|\underline Y})
\end{align}
for an arbitrary PDF ${\widehat f_{{\underline U}}}\equiv{\widehat
f_{{\underline U}}}(\underline u|\tau)$, where we  define two
functionals, $\mathcal{L}(\cdot,{\tau}):V \to \mathbb{R}$, where $V
\equiv \{\widehat f_{\underline U}: \mathbb{R}^N \to [0,1] \}$, and
the Kullback-Leibler (KL) divergence ${\text{KL}} ({\widehat
f_{{\underline U}}}||f_ {{\underline U}|\underline Y})$, as
\begin{align}\label{lower_bound_EM}
\begin{gathered} \mathcal{L}({\widehat
f_{{\underline U}}},{\tau}) \equiv {\mathbb{E}_{{{\widehat
f}_{{\underline U}}}}}[ \log f_{\underline Y, {\underline U}}(\underline y,{\underline u}|{\tau})]
+ \mathbb{H}({\widehat f_{{\underline U}}}) \hfill \\
\end{gathered},\\
\begin{gathered} {\text{KL}}({\widehat
f_{{\underline U}}}||{f_{{\underline U}|\underline Y}}) \equiv \mathbb{E}_{{\widehat
f}_{{\underline U}}}\left[ \log  {\frac{{\widehat f_{{\underline U}}({\underline u}|\tau)}}{{f_{{\underline U}|\underline Y}({\underline u}|\underline y,{\tau})}}}  \right] \hfill \\
\end{gathered}.
\end{align}
Note in \eqref{EM_LH} that the lower bound, \emph{i.e.}, $\log
{f_{\underline Y}}(\underline y|{\tau})\geq\mathcal{L}({\widehat
f_{{\underline U}}},{\tau})$, holds true since the  the KL
divergence is non-negative.

The EM algorithm consists of two-stages for iteratively maximizing
the log-likelihood \eqref{EM_LH}. Let ${\tau^t}\equiv\{q,\sigma_0^2\}^t$ denote the
current estimate of the parameter set. Then, we derive the EM update for the next  estimate ${\tau^{t+1}}\equiv\{q,\sigma_0^2\}^{t+1}$ as follows.

\begin{table*}
\small
\renewcommand{\arraystretch}{1.2}
\caption{List of recent solvers in the performance validation}
\label{algotable}
 \centering
\begin{tabular}{||c||c|c|c||}
\hline
Solvers &  Optimization Setup                        &  Parameter Tuning    &  Solver Type \\
\hline \hline
ssAMP-BGFD   & MMSE + BG prior                       &  $q,\sigma_0,\Delta$ (Oracle/EM)      & Sum-product AMP   \\
 \hline
GrAMPA-BG  \cite{GrAMPA}  &  MMSE + BG prior         & $q,\sigma_0,\Delta$ (Oracle)         &  Sum-product AMP (GAMP-based)  \\
\hline
TVAMP    \cite{TV_AMP}   &    MAP + Laplacian prior  & $\lambda$ (Empirically optimal)     &  Max-sum AMP + FLSA \cite{FLSA} or Condat's 1DTV \cite{condat}  \\
\hline
EFLA     \cite{sfa}     &     TV method \eqref{TV}   & $\lambda$ (Empirically optimal)        & First-Order + FLSA \cite{FLSA}   \\
\hline TV-CP \cite{CP}  &   TV method \eqref{TV}     & $\lambda$ (Empirically optimal)        &First-Order  + Chambolle-Pock \cite{CP}        \\
\hline
\end{tabular}
\end{table*}

\subsubsection{In the E-step} Given the current estimate ${\tau^t}$, we find the PDF ${\widehat f_{{\underline U}}}$
maximizing the lower bound $\mathcal{L}({\widehat
f_{{\underline U}}},{\tau^t})$. For the optimum, we obviously need to set $\widehat f_{\underline U} = f_{{\underline U}|\underline Y}({\underline u}|\underline y,{\tau}^t)$ such that
the KL divergence becomes zero and the log-likelihood achieves the lower bound, \emph{i.e.}, $\log {f_{\underline Y}}(\underline y|{\tau^t})=\mathcal{L}({\widehat f_{{\underline U}}},{\pi^t})$.
The optimum PDF $\widehat f_{\underline U}$ is obtained by the product of marginal posterior of $U_d$. Namely, we have
\begin{align}
\widehat f_{\underline U}=f_{{\underline U}|\underline Y}({\underline u}|\underline y,{\tau})=\prod\nolimits_{d = 1}^{N-1} {{f_{{U_d}|\underline Y}}({u_d}|\underline y,{\tau} )},
\end{align}
and then we find
\begin{align}\label{posterior_uk}
\begin{gathered}
  {f_{U_d |\underline Y }}(u_d|\underline y ,{\tau}) =\frac{{{f_{{U_d}}}({u_d}|\tau )\mathcal{N}(u_d;{\rho _{i_2}} - {\rho _{i_1}},2\theta) } }{{\int {{f_{{U_d}}}({u_d}|\tau )\mathcal{N}(u_d;{\rho _{i_2}} - {\rho _{i_1}},2\theta) d{u_d}} }}\hfill \\
  \,\,\,\,\,\,\,\,\,\,\,\,\,\,\,\,\,\,\,\,\,\,\,\,\,\,\,\,\,\,\,\,\,\,\,\,\,\,\,=(1 - {\pi _d})\delta ({u_d}) + {\pi _d}\mathcal{N}({u_d};{\gamma _d},\nu ),\hfill \\
\end{gathered}
\end{align}
with some parameters definitions:
\begin{align}
&\begin{gathered}
  {\pi _d} \equiv \frac{1}{{1 + {{{\frac{1-q}{{q}}\frac{ \mathcal{N}(0;{\rho _{i2}} - {\rho _{i1}},2\theta ) } {  \mathcal{N}({\rho _{i_2}} - {\rho _{i_1}};0,2\theta  + \sigma _0^2) }}}}}} \hfill \\
  \end{gathered},\\
 & \begin{gathered}
  {\gamma _d} \equiv \frac{{   {{\rho _{i_2}} - {\rho _{i_1}}}   }}{{\frac{2\theta}{{\sigma _0^2}} + 1 }} \hfill \\
    \end{gathered},\\
  &\begin{gathered}
  \nu  \equiv \frac{1}{{\frac{1}{{\sigma _0^2}} + \frac{1}{{2\theta }}}} \hfill \\
    \end{gathered},
\end{align}
for $i_1,i_2 \in ne(d)$, where we note that
$\rho_{i_1},\rho_{i_2},\theta$ are approximated by the ssAMP-BGFD
iteration of Algorithm \ref{algo2}.

\subsubsection{In the M-step} We fix the PDF ${\widehat f_{{\underline U}}}$ by the E-step and
maximize the lower bound $\mathcal{L}({\widehat
f_{{\underline  U}}},{\mathbf{\tau }^t})$ with respect to ${\mathbf{\tau }}$
to find an next estimate ${\mathbf{\tau }^{t+1}}$. Since the
entropy term is independent of $\mathbf{\tau}$ in
\eqref{lower_bound_EM}, this maximization clearly can be
\begin{align}\label{Mstep}
\begin{gathered}
{\tau^{t + 1}} = \arg \mathop {\max }\limits_\tau
{\mathbb{E}_{{{\widehat f}_{{\underline U}}}}}[\log f_{\underline Y,\underline  U}(\underline y, {\underline u}|\mathbf{\tau}^{t} )] \hfill \\
\,\,\,\,\,\,\,\,\,\,\,\,\,= \arg \mathop {\max }\limits_\tau \sum \limits_{d=1}^{N-1}
{\mathbb{E}_{{{\widehat f}_{{U_d}}}}}[\log  f_{U_d}(u_d |\mathbf{\tau}^{t} )] \hfill \\
\end{gathered}
\end{align}
where the equality for the second line holds since we can express the joint PDF as
$f_{\underline Y,\underline U}(\underline y, {\underline u}|\mathbf{\tau} )= C \times \prod\nolimits_{d = 1}^{N-1}  {f_{U_d}}(u_d|\tau )$ for a $\tau$-independent term
$C ={f_{\underline Y |{\underline X}}}(\underline y |{\underline x})$.

The M-step maximization \eqref{Mstep} need to be solved with respect
to each parameter of  $\tau$. We omit the detailed manipulation to
handle this M-step maximization by referring readers to the work of
Vila and Schniter (Section III-B of \cite{EM-BG-GAMP}). Finally, we
formulate our EM update  as:
\begin{align}
&\begin{gathered}\label{EMupdate1}
  {q^{t + 1}} = \frac{1}{{N - 1}}\sum\limits_{d=1}^{N-1 } {\pi _d^t}\hfill \\
  \end{gathered}\\
  &\begin{gathered}\label{EMupdate2}
  {(\sigma _0^2)^{t + 1}} = \frac{1}{{{q^{t + 1}}(N - 1)}}\sum\limits_{d=1}^{N-1 } {{\pi _d}\left( {| \gamma _d^t{|^2} + {\nu ^t}} \right)}.\hfill \\
    \end{gathered}
\end{align}
This EM-tuning routine can be optionally inserted at the end of the
ssAMP-BGFD iteration. With \eqref{Delta_appx}, \eqref{EMupdate1},
and \eqref{EMupdate2},  ssAMP-BGFD can be parameter-free.

\section{Performance Validation }
In this section, we validate performance of the ssAMP-BGFD algorithm
with extensive empirical results\footnote{ We inform that all
experiments here were performed by MATLAB Version: 8.2.0.701
(R2013b).}. Three types of experimental results will be discussed in
this section:
\begin{enumerate}
\item Noiseless phase transitions,
\item Normalized MSE (NMSE)
convergence over iterations,
\item Average CPU runtime.
\end{enumerate}
All these experimental results were averaged using the Monte Carlo
method with $100$ trials. At each Monte Carlo trial, we took a
synthetic measurement vector  $\underline y$ by realizing a signal
$\underline x_0$, and an AWGN vector $\underline w$.

\begin{figure*}
\centering
\includegraphics[width=16cm]{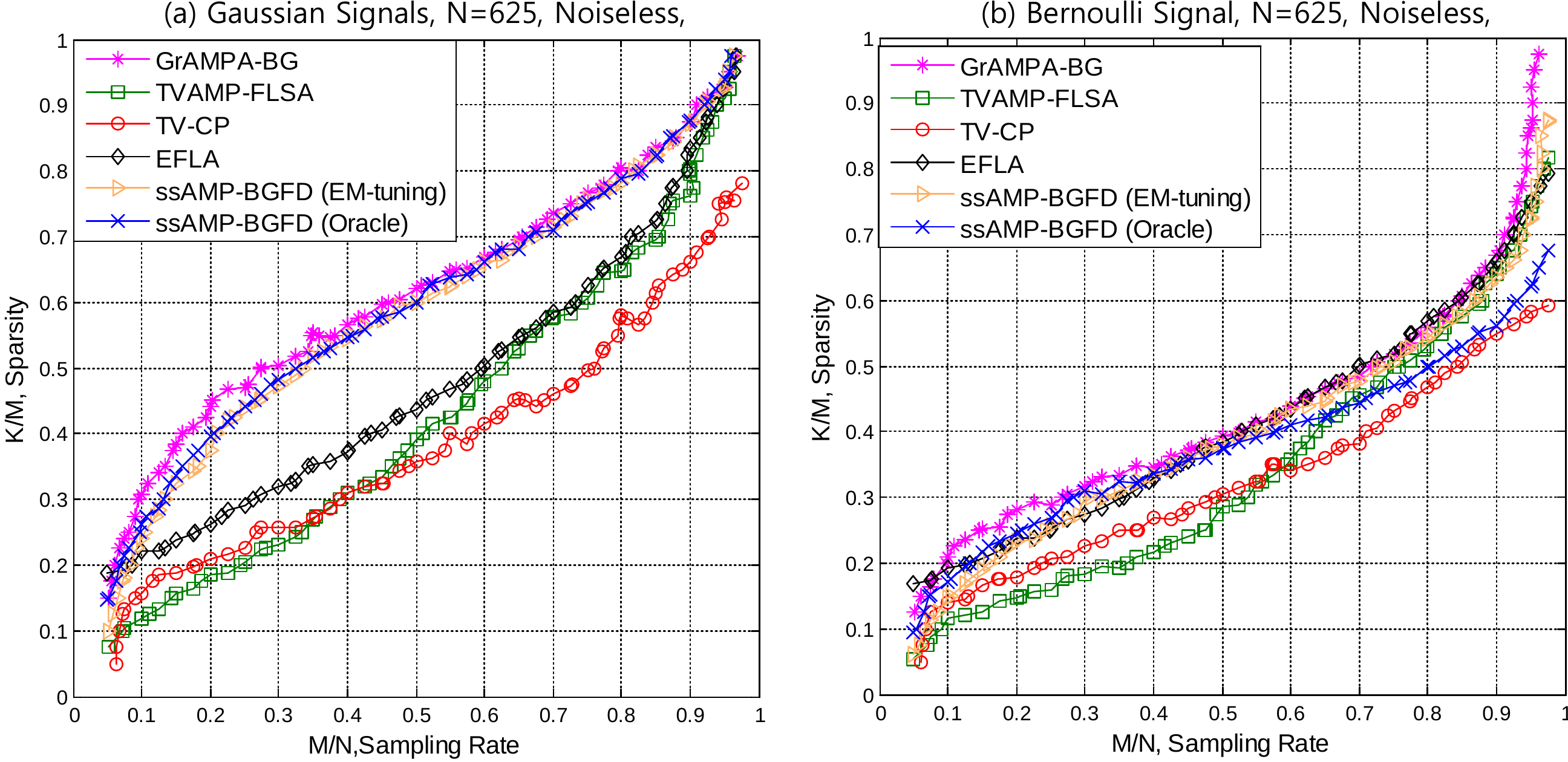}
\caption{Empirical  noiseless PT curves with std. Gaussian
$\mathbf{H}$ for two types of the 1D piecewise-constant (PWC)
signals: (a) Gaussian PWC signals and (b) Bernoulli PWC signals,
which are generated by the PDFs given in Table \ref{table2} where we
set the signal length $N=625$ and the variance $\sigma_0^2=1$.}
\label{fig_PTC}
\end{figure*}

In this experiment, we included recent solvers for the CS recovery
with 1D-FD sparsity, listed in Table \ref{algotable}, for a
comparison purpose. The source codes of each solver was basically
obtained from each authors's webpage\footnote{The source code of
EFLA is obtained from the SLEP 4.1 package \cite{SLEP}; The source
codes of GrAMPA was downloaded from
http://www2.ece.ohio-state.edu/$\sim$schniter/GrAMPA
(gampmatlab20141001.zip); The source codes  of ssAMP-BGFD is from
https://sites.google.com/site/jwkang10/home/ssamp.}, but TV-CP and
TVAMP were implemented by the authors. We provide two version of
ssAMP-BGFD according to its EM option for the prior parameter
learning. TVAMP was implemented in two ways: ``TVAMP-FLSA" and
``TVAMP-Condat" according to the denoiser implementation of
\eqref{etaTV}\footnote{TVAMP-FLSA is based on the \emph{Fused lasso
signal approximator} (FLSA) implementation \cite{FLSA}, and
TVAMP-Condat is based on the recent direct 1D-TV implementation
\cite{condat}.}. In addition,  we configure GrAMPA to use the BG
prior in this experiment, referring to the solver as ``GrAMPA-BG" to
specify its prior attribute.

We note that ssAMP-BGFD and GrAMPA-BG were configured by the
oracle-tuning for the prior parameter $q,\sigma_0$ and the noise
variance $\Delta$, but ssAMP-BGFD could be parameter-free with the
EM-tuning (discussed in Section III-D). For TV-AMP, EFLA, and TV-CP,
we used an empirically optimal $\lambda$ for each
($\frac{K}{M},\frac{M}{N}$). Finally, we set the initial guess of
the signal estimate to a zero vector for all the solvers. For your
information, we note that the empirical results, reported in this
paper, have some changes from the results given in our conference
paper \cite{ssAMP1} due to some mis-configuration corrections.

\subsection{Noiseless Phase Transition}
\subsubsection{Experimental setup}
For each  PT curve, we basically considered a $ 38 \times 38$ grid
where we uniformly divided the range $\frac{M}{N} \in [0.05,0.99]$
as the x-axis and the range $\frac{K}{M} \in [0.05,0.99]$ as the
y-axis with the stepsize $0.025$. A PT curve is connection of
experimental points having 0.5 success rate of the signal recovery,
where the recovery success is declared when $\text{NMSE} \equiv
\frac{{||{{\underline x }_0} - \underline {\widehat x}
||_2^2}}{{||{{\underline x }_0}||_2^2}} \leq 10^{-4}$. We set the
number of maximum iterations to $t^*=2000$, and the iteration
stopping tolerance was very tightly set to $ \frac{{||{{\underline
\mu }^{(t)}} - {{\underline \mu }^{(t +
1)}}||_2^{2}}}{{||{{\underline \mu }^{(t)}}||_2^2}} \leq
\text{tol}=10^{-14}$; hence, the PT curves are supposed to represent
algorithm performance after convergence.

\subsubsection{Comparison over the other solvers}
In Fig.\ref{fig_PTC}, we provide a PT comparison over the recent
solvers listed in Table \ref{algotable}. For this, we fixed
$\mathbf{H}$ to the standard Gaussian matrix whose entries are drawn
from $\mathcal{N}(h_{ji};0,\frac{1}{M})$, setting to $N=625$. Then,
we draw PT curves for two types of the signal statistics given in
Table \ref{table2}.

In the Gaussian case, we observe from Fig.\ref{fig_PTC}-(a) that
GrAMPA-BG provides the state-of-the-art, and ssAMP-BGFD  retains its
place very close to GrAMPA-BG. Those two algorithms significantly
improve on the PT performance of the others because their BG prior
has a very good match with the Gaussian statistics. We also note in
the Gaussian case that the EM method exactly tunes the prior
parameters of ssAMP-BGFD such that its PT curve coincides with that
by the oracle tuning.

For the Bernoulli case, Fig.\ref{fig_PTC}-(b) reports that
ssAMP-BGFD using EM is the closest to GrAMPA-BG together with EFLA,
and better than TV-CP and TVAMP-FLSA even though its advantage is
less remarkable compared to the Gaussian case. In this case, the
oracle tuning of ssAMP-BGFD is not as fine as in the Gaussian case
because the BG prior is not basically able to provide an accurate
description to the statistic of the Bernoulli PWC.

\begin{figure}
\centering
\includegraphics[width=8.5cm]{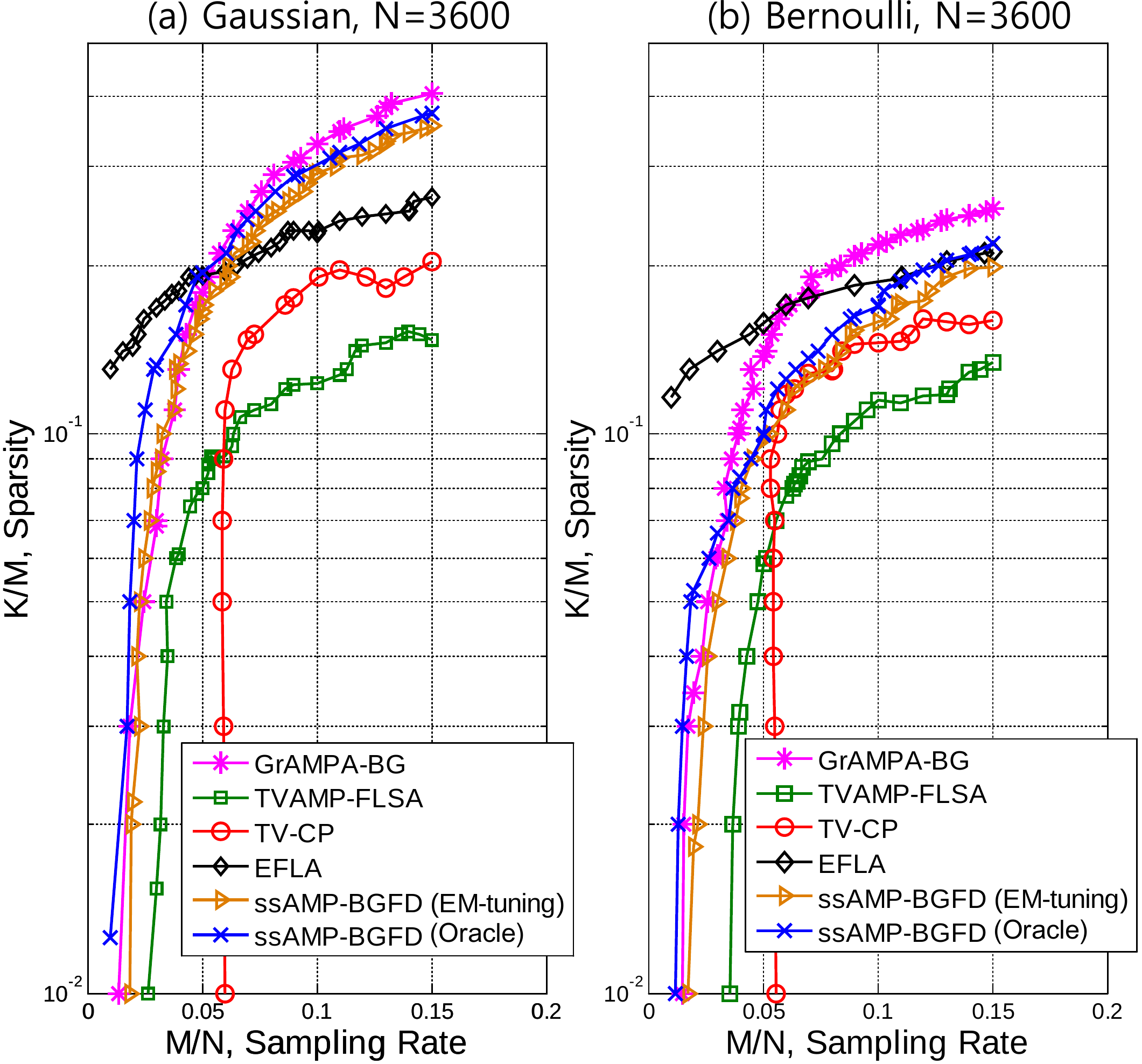}
\caption{Empirical  noiseless PT curves with std. Gaussian
$\mathbf{H}$ for the case of $M \ll N$ with $N=3600$.}
\label{fig_PTC2}
\end{figure}

To better understand the PT characteristic when small $M/N$ ($M \ll
N$), we fixed $N=3600$ and constructed a $22 \times 8$ uniform grid
of ($\frac{K}{M},\frac{M}{N}$) with the stepsize $0.02$ where the
range of the x-axis is $\frac{M}{N} \in [0.01,0.15]$, and the range
of the y-axis is $\frac{K}{M} \in [0.01,0.43]$. We observe from
Fig.\ref{fig_PTC2} that as $M/N \to 0$, the PT curves of ssAMP-BGFD
and GrAMPA-BG becomes nearly identical, worse than that of EFLA, and
much better than those of TVAMP-FLSA and TV-CP.

These comparison results support that ssAMP-BGFD shows the PT
performance closely approaching the state-of-the-art by GrAMPA-BG,
being superior to the others.

\begin{table}
\renewcommand{\arraystretch}{1.1}
\small
 \centering
 \caption{Statistics of the 1D-PWC signals $\underline X$} \label{table2}
\begin{tabular}{||c|c||}
\hline \hline
Type & Signal PDFs,  $f_{U_d}(u_d)$ \\
\hline Gaussian PWC&  $(1 - q)\delta(u_d) + q\mathcal{N}(u_d;0,\sigma_0^2 )$            \\
\hline Bernoulli PWC&  $(1 - q)\delta(u_d) + q\mathcal{U} (u_d \in \{  - {\sigma _0},{\sigma _0}\} ;\frac{1}{2})$               \\
\hline \hline
\end{tabular}
\end{table}

\begin{figure*}
\centering
\includegraphics[width=18cm]{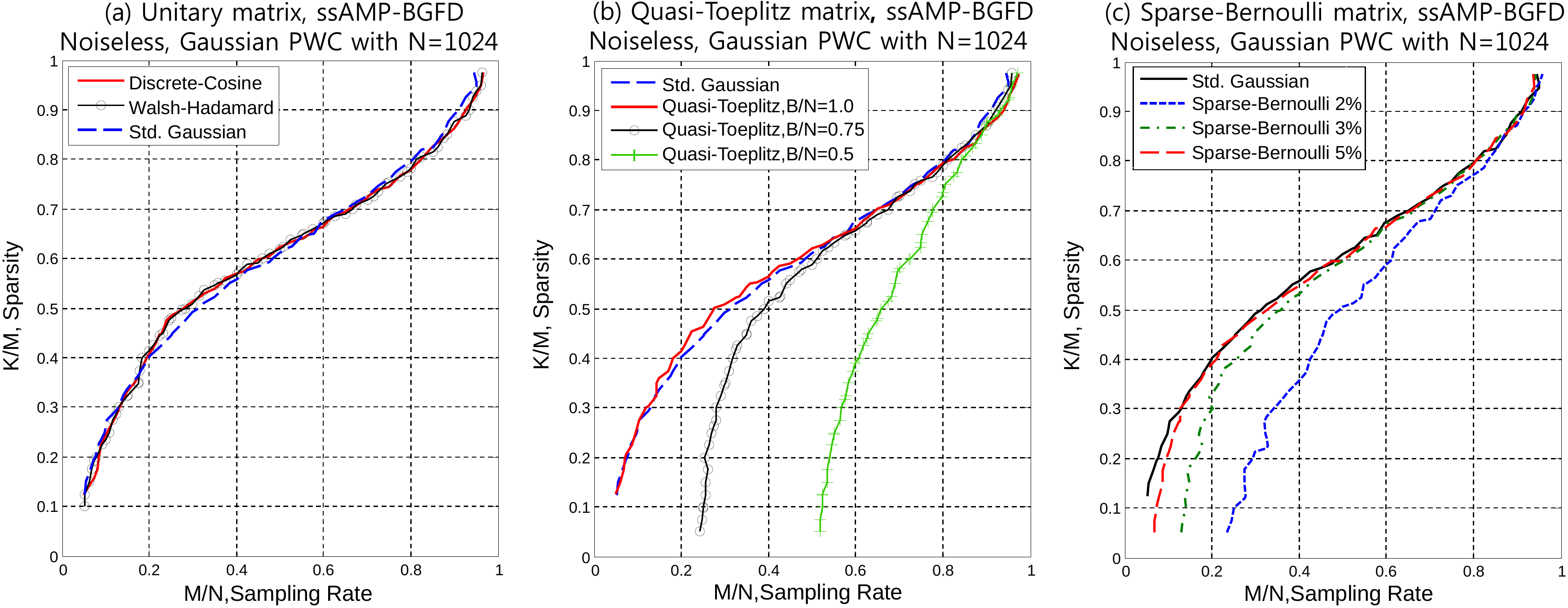}
\caption{Empirical noiseless PT curves of  ssAMP-BGFD (with the
EM-tuning) working with non-i.i.d.-random  $\mathbf{H}$ and
column-sign-randomization: (a) sub-sampled unitary matrices (DCT and
WHT), (b) quasi-Toeplitz matrices with the damping factor
$\beta=0.5$, (c) sparse-Bernoulli matrices for a variety of the
matrix sparsity, where we consider Gaussian PWC signal with
$N=1024$.} \label{fig_PTC3}
\end{figure*}

\subsubsection{PT curve of ssAMP-BGFD with RIP matrices $\mathbf{H}$}
Candes \emph{et al.} discussed a natural property on the measurement
matrix $\mathbf{H}$ (abbreviated by D-RIP), which is a variant of
the restricted isometry property (RIP) for the analysis CS setup
\cite{candes}. Then, they also stated using the result of
\cite{krahmer} that any $\mathbf{H}$ satisfying the standard RIP
requirement, will also satisfy the D-RIP with
``column-sign-randomization". Specifically, instead of
\eqref{system}, we consider the measurement generation:
\begin{align}\label{csr}
\underline Y  = \underbrace {\mathbf{H}}_{{\text{RIP
matrix}}}{\text{diag}}(\underbrace {[1,1, - 1,1..., - 1]}_{N{\text{
independent random signs}}})\,\underline X  + \underline W.
\end{align}
This leads us to test practical RIP matrices for the proposed
algorithm, such as unitary matrices, quasi-Toepliz matrices and/or
deterministic matrices, which aims to overcome the practical
limitation of AMP (discussed in Section II). We refer the reader to
\cite{krahmer} for specific RIP condition of each matrix listed
above. In this experiment, we demonstrate that ssAMP-BGFD works well
with such RIP matrices and the column-sign-randomization by showing
empirical evidences for the Gaussian PWC signals with $N=1024$.

\begin{figure}[!t]
\centering
\includegraphics[width=8.8cm]{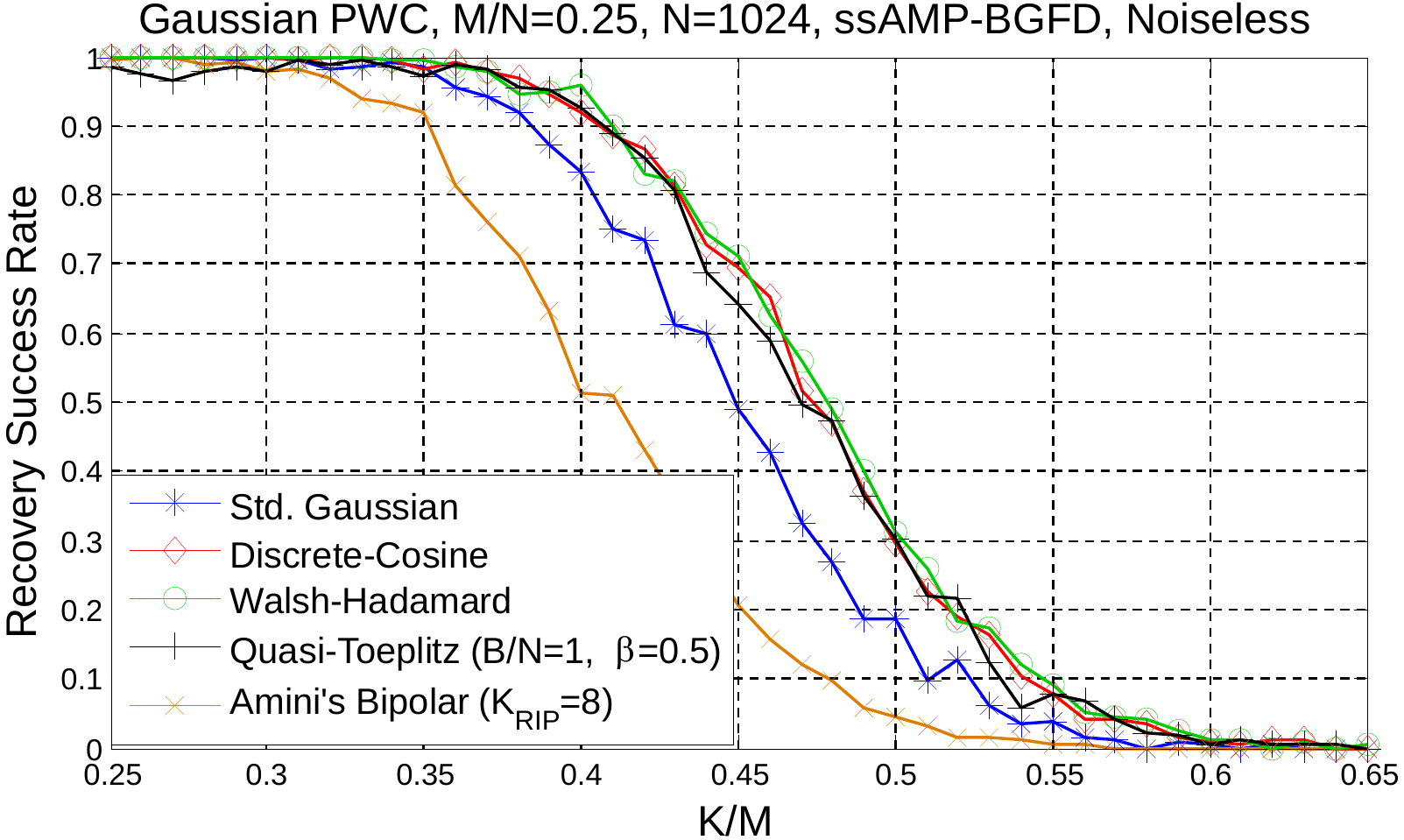}
\caption{Empirical PT  comparison of ssAMP-BGFD (EM-tuning) with
various types of the matrices $\mathbf{H}$: Amini's deterministic
bipolar matrix ($K_{\text{RIP}}=8$) \cite{Amini}, DCT and WHT
matrices, a quasi-Toeplitz matrix $(B/N=1.0)$ with damping
($\beta=0.5$), and the std. Gaussian matrix, where we consider
$M/N=0.25$ and Gaussian PWC with $N=1024$. } \label{fig_PTC5}
\end{figure}

\begin{itemize}
\item {\bf{With sub-sampled  unitary matrices:}} We test the two
unitary systems:  \emph{Discrete Cosine Transform} (DCT) and
\emph{Walsh-Hadamard Transform} (WHT) with ssAMP-BGFD. We construct
$\mathbf{H}$ by randomly sampling $M$ rows from the $N \times N$ DCT
or WHT matrix. For such matrices $\mathbf{H}$, the complexity of the
matrix-vector multiplication can be reduced to $\mathcal{O}(N \log
N)$ from $\mathcal{O}(MN)$ via the fast DCT/WHT method.
Fig.\ref{fig_PTC3}-(a) shows that the PT curve of the DCT and WHT
matrices coincides with that of the standard Gaussian matrix.
\item {\bf{With quasi-Toeplitz matrices with damping:}} We
consider  quasi-Toeplitz  $\mathbf{H}$ for ssAMP-BGFD: the first row
consists of $B$ zero-mean Gaussian coefficients, and each row of
$\mathbf{H}$ is a copy of the first row with cyclic permutation
\cite{Bajwa},\cite{rand_filter}. This matrix requires memory storage
only for the $B$ random numbers, enabling  fast matrix-vector
multiplications using the FFT method ($\mathcal{O}(N \log N)$
complexity). In addition, the row sampling of $\mathbf{H}$ need not
be random in contrast to the unitary case. On the other hand, in
this case, the columns of $\mathbf{H}$ are severely correlated, and
it may lead to the AMP divergence. For this, we use a simple damping
method to stabilize the ssAMP-BGFD iteration. Namely,  for the
residual update, we use
\begin{align}
\small
\begin{array}{l}
\small {\underline r ^{(t)}} = (1 - \beta ){\underline r ^{(t - 1)}}
+ \beta \left( {\underline y  - {\mathbf{H}}{{\underline \mu
}^{(t)}} + {{\underline r }^{(t - 1)}}\frac{N}{M}\left\langle {\eta
'(\cdot )} \right\rangle }\right)
\end{array}\nonumber
\end{align}
where $0< \beta \leq 1$ is the damping factor. As shown in
Fig.\ref{fig_PTC3}-(b), we test the quasi-Toeplitz matrices for
three cases , $B/N=0.5$, $B/N=0.75$ and $B/N=1.0$ with the damping
factors $\beta=0.5$.
\item{\bf{With deterministic
matrices:}} Several deterministic construction of $\mathbf{H}$ have
been developed to overcome some drawbacks of the random $\mathbf{H}$
\cite{DeVore},\cite{Amini}: mainly, there are no efficient methods
to verify whether a specific realization of the random $\mathbf{H}$
meets the RIP requirement. DeVore provided a deterministic
construction of cyclic binary $\mathbf{H}\in \{0,1\}^{M \times N}$
satisfying the RIP under some conditions \cite{DeVore}. Then, Amini
\emph{et al.} made a connection between the DeVore's approach and
channel coding theory (specifically BCH codes) and suggesting
construction of cyclic bipolar $\mathbf{H}\in \{-1,1\}^{M \times N}$
\cite{Amini}. One disadvantage of such deterministic $\mathbf{H}$ is
that the matrix size $(M,N)$  is restricted by the code length.
Under the ssAMP-BGFD recovery, we compare a PT curves by the $255
\times 1024$ Amini's matrix satisfying RIP order of
$K_{\text{RIP}}=8$, to PT curves by the $256 \times 1024$ matrices
considered above. Fig.\ref{fig_PTC5} shows that Amini's matrix works
well with ssAMP-BGFD even through its PT curve slightly
underperforms the PT curve by the other matrices.
\end{itemize}

\subsubsection{PT curve of ssAMP-BGFD  with matrix sparsity}
We consider the use of  sparse matrices with ssAMP-BGFD. This is
motivated by \emph{Low-Density Parity-Check} (LDPC) codes as the
works in \cite{BHT-BP}-\cite{SuPrEM},\cite{LDPC-CS}.  The use of the
sparse $\mathbf{H}$  provides an  accelerated fast matrix-vector
multiplication method (its complexity is proportional to the number
of nonzeros in $\mathbf{H}$), requiring small memory to store the
matrix entries. We examine sparse-Bernoulli random $\mathbf{H} \in
\{0,-1,1\}^{M \times N}$ whose column weight is fixed to $L$ such
that the matrix sparsity is $L/M \times 100$. Fig.\ref{fig_PTC3}-(c)
reports the corresponding PT curves for a variety of the matrix
sparsity: 2,3, and 5\% sparsity.

\begin{figure}[!t]
\centering
\includegraphics[width=8.7cm]
{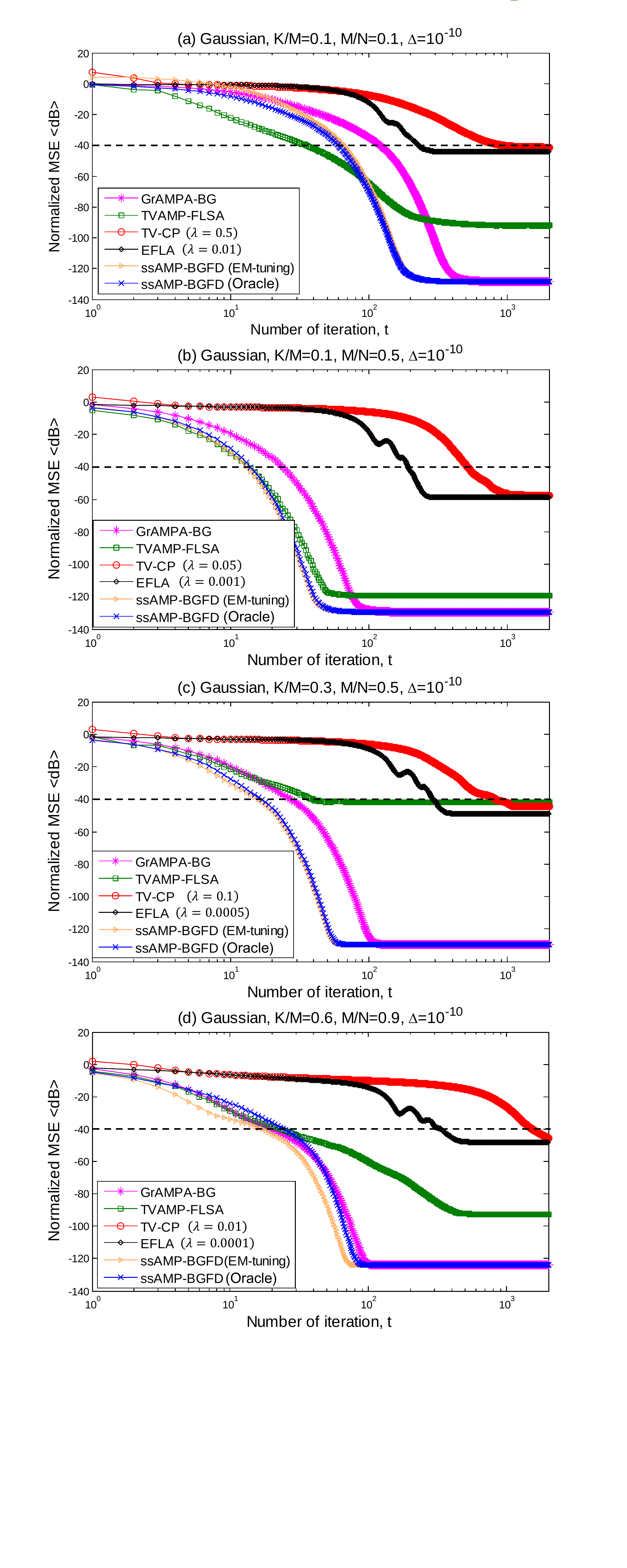}
\caption{NMSE convergence over iterations for the four cases of
($\frac{K}{M},\frac{M}{N}$) where Gaussian PWC signals are with
$N=3600$ and $\sigma_0^2=1$; the noise variance is set to
$\Delta=10^{-10}$; the matrix $\mathbf{H}$ is set to the std.
Gaussian.} \label{fig13}
\end{figure}

\subsection{NMSE Convergence over Iterations}
\subsubsection{Experimental setup}
We measured NMSE over iterations  for the four different cases of
($\frac{K}{M},\frac{M}{N}$):
\begin{itemize}
\item  Case (a) - $\frac{K}{M}=0.1,
\frac{M}{N}=0.1$,
\item  Case (b) - $\frac{K}{M}=0.1,
\frac{M}{N}=0.5$,
\item  Case (c) - $\frac{K}{M}=0.3,
\frac{M}{N}=0.5$,
\item Case (d) - $\frac{K}{M}=0.6,
\frac{M}{N}=0.9$,
\end{itemize}
which are points satisfying the Gaussian PT curve of all the solvers
(see Fig.\ref{fig_PTC}-(a)). In this experiment, we set the noise
variance to $\Delta=10^{-10}$, the signal length to $N=3600$, and
consider the standard Gaussian $\mathbf{H}$. Also, we inform that
all the solvers were set to run by $t^*=2000$ iterations without any
stopping criterion.

\subsubsection{Discussion for the Gaussian PWC case}
\begin{figure}[!t]
\centering
\includegraphics[width=7.5cm]{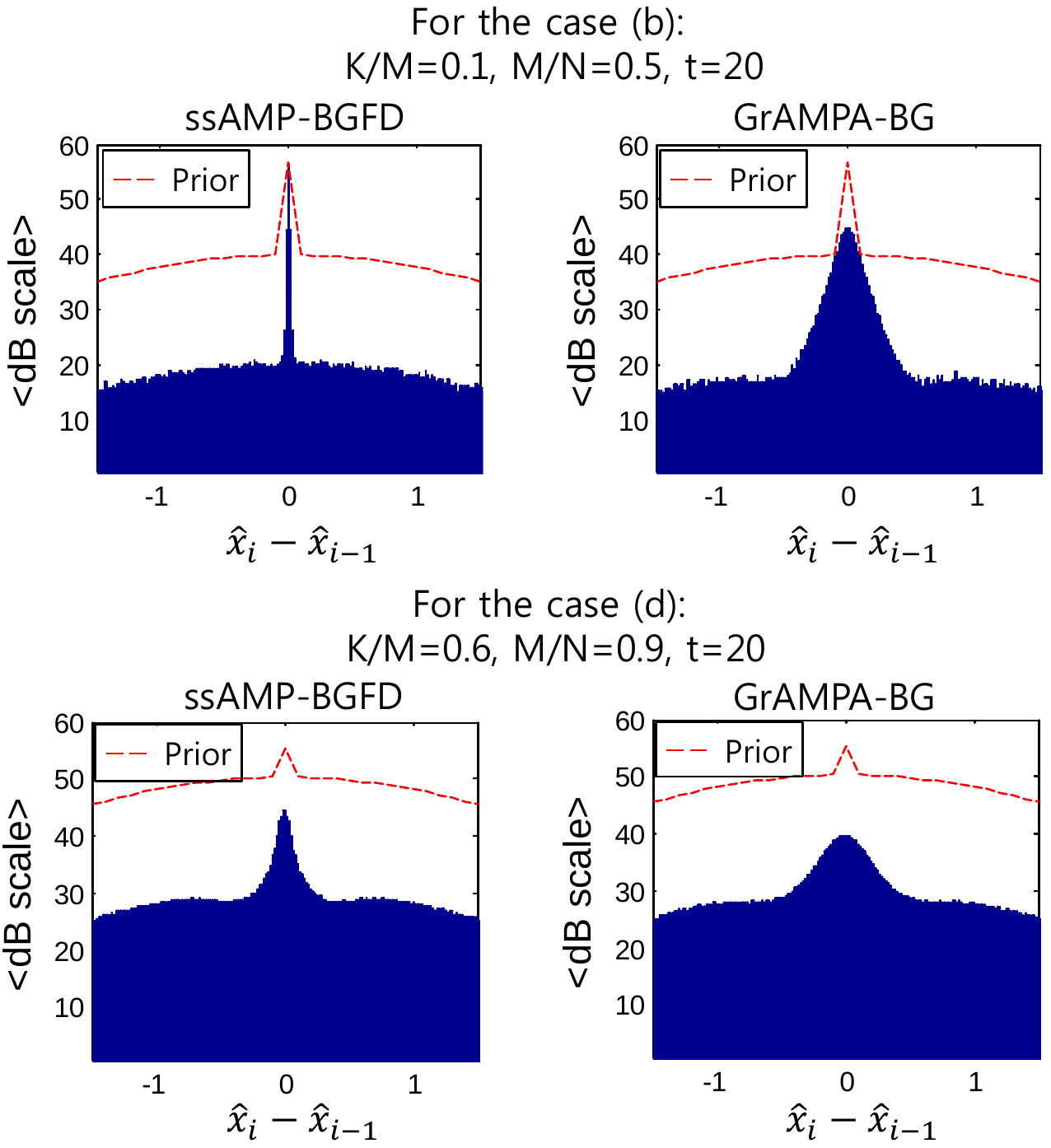}
\caption{Empirical PDFs  (in dB scale) of the estimated Gaussian PWC
by ssAMP-BGFD (w/ the EM-tuning) and GrAMPA-BG. These data were
obtained with $N=10^{5}$, $\sigma_0^2=1$, $\Delta=10^{-10}$, and
iteration $t=20$ for the two cases: Case (b) ($\frac{K}{M}=0.1$,
$\frac{M}{N}=0.5$) and Case (d) ($\frac{K}{M}=0.6$,
$\frac{M}{N}=0.9$)} \label{fig_hist}
\end{figure}

We consider the Gaussian PWC case first. Table \ref{table6} and
Fig.\ref{fig13} reports that in all the cases, ssAMP-BGFD converges
remarkably faster than EFLA and TV-CP, being advantageous over
TVAMP-FLSA and GrAMPA-BG. Although TVAMP-FLSA shows the fastest
convergence rate in the case (a), it pales into insignificance  due
to an non-negligible NMSE gap from ssAMP-BGFD at the fixed-point. In
such a aspect, GrAMPA-BG is the most comparable, but there exists an
uniform gap between ssAMP-BGFD and GrAMPA-BG in convergence rate. We
state that this gap is caused by difference of  the sF2V message
modeling methods (discussed in Section III-C). We support our
statement by plotting empirical PDFs of the estimated Gaussian PWC
at iteration $t=20$, as shown in Fig.\ref{fig_hist}. In case (b),
GrAMPA-BG's method induces approximation errors in the sF2V
modeling, delaying its convergence, resulting in an empirical PDF
with a blunt peak at $t=20$. In contrast, ssAMP-BGFD's method does
not cause such errors, promoting its convergence, showing a sharp
PDF whose peak nearly coincides with that of the prior PDF even at
$t=20$. We also note from Fig.\ref{fig_hist}   that in the case (d),
ssAMP-BGFD requires far more iteration  than $t=20$ for its
convergence, implicating that its convergence advantage is decayed
compared to the case (b). This is because the BG-based modeling
method of ssAMP-BGFD, given in \eqref{pp_sf2v}, less effective for
non-sparse signals having  high $K/M$.

\subsubsection{Discussion for the Bernoulli PWC case}
In the Bernoulli PWC case, every algorithm basically requires more
iterations than the Gaussian case. In addition, we observe from
Table \ref{table7} that in the case (d), the oracle ssAMP-BGFD does
not achieve the NMSE =-40dB whereas ssAMP-BGFD with EM does. This
observation implicates that the EM-tuning effectively assists
ssAMP-BGFD to estimate the Bernoulli PWC signals using the BG prior.
This is also connected to the PT improvement of ssAMP-BGFD in
Fig.\ref{fig_PTC}-(b).

\begin{table}
\renewcommand{\arraystretch}{1}
\caption{The average number of iterations for achieving the
normalized MSE = -40 dB where $N=3600, \Delta=10^{-10}$, and
Gaussian PWC.} \label{table7}
 \centering
 \scriptsize
\begin{tabular}{||c||c|c|c|c||}
\hline\hline
                               & Case (a)  & Case (b)   & Case (c) &  Case (d)\\
Algorithms                     & K/M=0.1,  & K/M=0.1,  & K/M=0.3,  & K/M=0.6,  \\
                               &   M/N=0.1 &  M/N=0.5  &  M/N=0.5 & M/N=0.9 \\
\hline
\hline  ssAMP-BGFD                & 60             & $\mathbf{14}$     &17                 & 25\\
(Oracle)&&&&\\
\hline  ssAMP-BGFD                & 66               & $\mathbf{14}$     & $\mathbf{16}$     & $\mathbf{17}$ \\
(EM-tuning)&&&&\\
\hline        TV-CP          & 921              & 515               & 823               & 1525 \\
\hline        EFLA           & 228              & 192               & 289               & 335\\
\hline        TVAMP-FLSA     & $\mathbf{36}$    & $\mathbf{14}$      & 39               & 23\\
\hline        GrAMPA-BG      & 121              & 24                & 28                 & 20\\
\hline\hline
\end{tabular}
\bigskip
\renewcommand{\arraystretch}{1}
\caption{The average number of iterations for achieving the
normalized MSE = -40 dB where $N=3600, \Delta=10^{-10}$, and
Bernoulli PWC.} \label{table6}
 \centering
 \scriptsize
\begin{tabular}{||c||c|c|c|c||}
\hline\hline
                               & Case (a)  & Case (b)   & Case (c) &  Case (d)\\
Algorithms                     & K/M=0.1,  & K/M=0.1,  & K/M=0.3,  & K/M=0.6,  \\
                               &   M/N=0.1 &  M/N=0.5  &  M/N=0.5 & M/N=0.9 \\
\hline
\hline  ssAMP-BGFD      & 66             & $16$              & $\mathbf{25}$        & $\infty$   \\
(Oracle)&&&&\\
\hline  ssAMP-BGFD    & 75               & $\mathbf{14}$     & 28                   & $\mathbf{44}$  \\
(EM-tuning)&&&&\\
\hline        TV-CP          & 1044             & 570               & $\infty$             & $\infty$    \\
\hline        EFLA           & 234              & 191               & 324                  & 494         \\
\hline        TVAMP-FLSA     & $\mathbf{54}$    & 15                & $\infty$             & 69         \\
\hline        GrAMPA-BG      & 143             & 25                & 39                    &    47    \\
\hline\hline
\end{tabular}
\end{table}

\subsection{Average CPU Runtime}
In order to clarify the computational advantage of ssAMP-BGFD, we
provide a comparison of CPU runtime over the algorithms of Table
\ref{algotable}.
\subsubsection{Experimental setup}
In this experiment, we again considered the four cases of
($\frac{K}{M},\frac{M}{N}$) given in Section IV-B, the Gaussian PWC,
the noise variance $\Delta=10^{-10}$, and the standard Gaussian
$\mathbf{H}$. For a fair comparison, we set a target MSE since some
algorithms may run longer but give a better MSE without any stopping
criterion. Namely, we made all the algorithms to stop their
iterations when reaching the target MSE $\frac{{||{{\underline x
}_0} - \underline {\widehat x} ||_2^2}}{{||{{\underline x
}_0}||_2^2}} \leq 10^{-4}$ ($-40$ dB of NMSE). We loosely set the
maximum iterations to $t^*=2000$ based on the result of Section
IV-B. In addition, we only counted the cases where all the algorithm
achieve the target MSE. We inform that this runtime measuring was
performed by using the ``tic-and-toc" functions of MATLAB R2013b
with Intel Core i7-3770 CPU (3.40 GHz) and RAM 24GB. Finally, we
clarify that the 1D-FD matrix $\mathbf{D}$ for GrAMPA-BG and TV-CP
is declared by the ``sparse" attribute in MATLAB.


\subsubsection{Discussion}
\begin{figure}[!t]
\centering
\includegraphics[width=8.9cm]{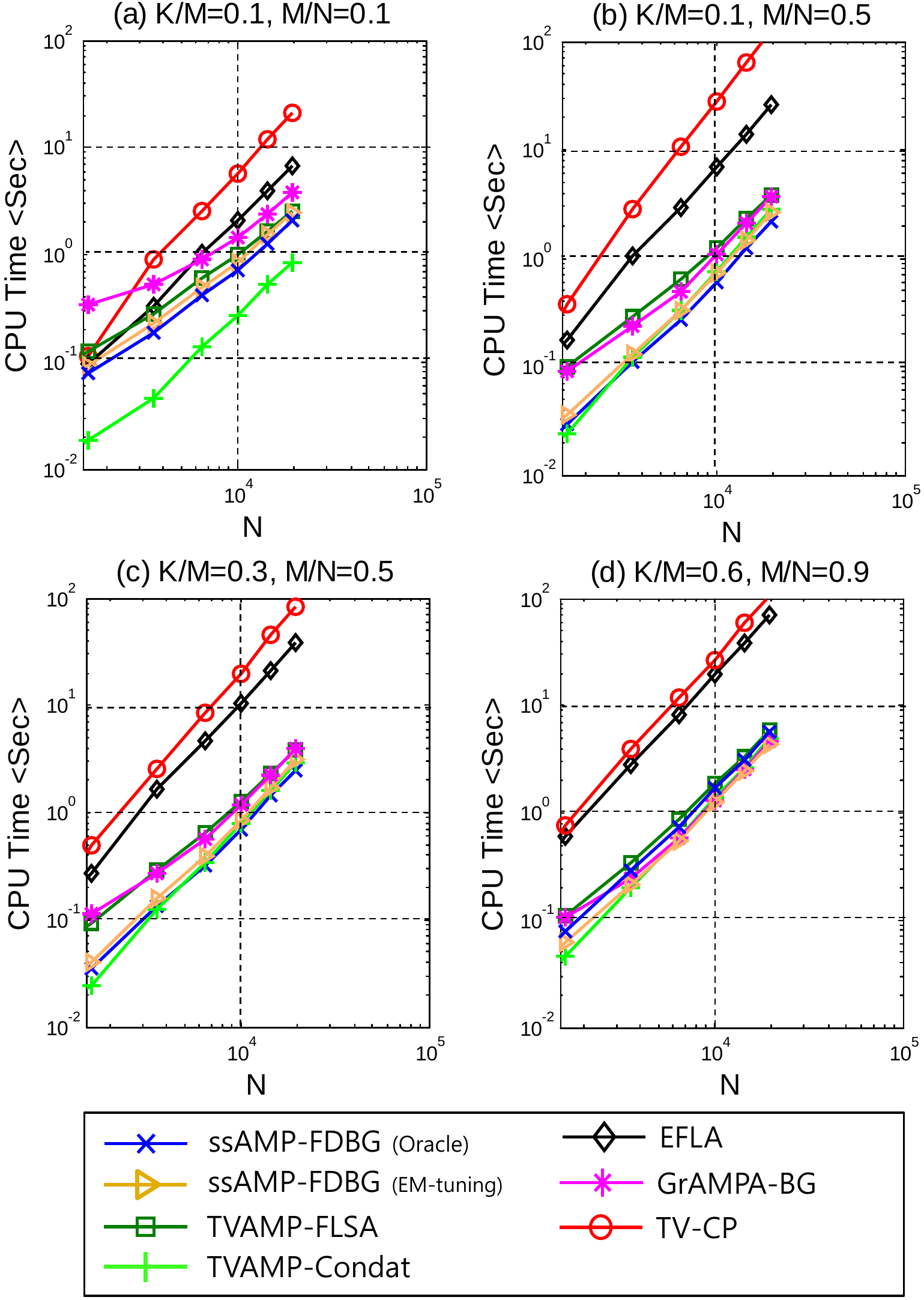}
\caption{CPU runtime comparison (in seconds) over signal length $N$
for the four cases of ($\frac{K}{M}$, $\frac{M}{N}$), where we
consider the Gaussian PWC signals $\underline X$, the noise variance
$\Delta=10^{-10}$, and target MSE $\frac{{||{{\underline x }_0} -
\underline {\widehat x} ||_2^2}}{{||{{\underline x }_0}||_2^2}} \leq
10^{-4}$. The matrix $\mathbf{H}$ is set to the std. Gaussian. We
used MATLAB with Intel Core i7-3770 CPU (3.40 GHz) with RAM 24GB for
this experiment. } \label{fig9}
\end{figure}

The complexity cost of the algorithms are  dominated by the
matrix-vector multiplications, \emph{i.e.}, $\mathbf{H}\underline
\mu$ and $\mathbf{H}^T\underline r$. In the case of ssAMP-BGFD and
TVAMPs, their per-iteration cost is straightforwardly
$\mathcal{O}(MN)$ since they include $\mathbf{H}\underline \mu$ and
$\mathbf{H}^T\underline r$ once in a lap of the iteration. EFLA is
also $\mathcal{O}(MN)$ by including variably 2$\sim$3 times of the
multiplications per iteration. For TV-CP and GrAMPA-BG, their cost
is naively $\mathcal{O}(N^2)$ due to the size of the 1D-FD operator
$\mathbf{D}\in \mathbb{R}^{N \times N-1}$, but the cost can be
reduced to $\mathcal{O}(MN)$ by applying a fast sparse matrix
multiplication method\footnote{MATLAB automatically supports the
fast multiplication method for matrices declared by ``sparse"
attribute \cite{sparse}.}. In Fig.\ref{fig9}, we take notice slopes
of the runtime curves which manifest the complexity cost. Since  the
rate $M/N$ is fixed for each case, the per-iteration cost
$\mathcal{O}(MN)$ becomes $\mathcal{O}(N^2)$; hence all the curves
approximately have slope `2' with sufficiently large $N$ when the
x-axis plot the length $N$ on a logarithmic scale.

\begin{table*}
\renewcommand{\arraystretch}{1.1}
\caption{Average CPU runtime (in seconds) per iteration $(N=10000,
\Delta=10^{-10})$, Gaussian PWC signals (by MATLAB R2013b with Intel
Core i7-3770 CPU (3.40 GHz) with RAM 24GB)} \label{table5}
 \centering
 \footnotesize
\begin{tabular}{||c||c|c|c|c||}
\hline\hline
Algorithms                     & (a) K/M=0.1,  M/N=0.1 & (b)  K/M=0.1, M/N=0.5  & (c) K/M=0.3,  M/N=0.5 & (d) K/M=0.6, M/N=0.9 \\
\hline
\hline  ssAMP-BGFD (Oracle)  &0.014$\pm$4.9e-4      & 0.052 $\pm$ 1.3e-3      & 0.050$\pm$1.1e-3       & 0.094$\pm$3.2e-3\\
\hline {\bf{ ssAMP-BGFD (EM-tuning)}}  & {\bf{0.012$\pm$5.4e-4}}       & {\bf{0.050 $\pm$ 2.0e-3 }}     &{\bf{0.049$\pm$2.2e-3}}        & {\bf{0.090$\pm$2.7e-2}}\\
\hline        TV-CP            & 0.011$\pm$4.5e-4       & 0.049 $\pm$ 7.0e-4      & 0.049$\pm$7.1e-4       & 0.088$\pm$1.8e-2\\
\hline        EFLA             & 0.017$\pm$2.8e-3       & 0.063 $\pm$ 9.6e-3      & 0.062$\pm$9.7e-3       & 0.128$\pm$5.1e-2\\
\hline        TVAMP-FLSA       & 0.027$\pm$1.3e-3       & 0.055 $\pm$ 2.0e-3      & 0.054$\pm$1.8e-3       & 0.081$\pm$2.0e-3\\
\hline      {\bf{   TVAMP-Condat }}    & {\bf{0.010$\pm$3.3e-4}}       & {\bf{0.038 $\pm$ 1.2e-3}}      & {\bf{0.038$\pm$1.0e-3}}       & {\bf{0.066$\pm$1.6e-3}}\\
\hline        GrAMPA-BG        & 0.012$\pm$4.8e-4       & 0.051 $\pm$ 3.8e-3      &0.052$\pm$5.4e-3        & 0.092$\pm$3.8e-2\\
\hline\hline
\end{tabular}
\end{table*}

Then, what are the factors  distinguishing the superiority of the
runtime curves in this comparison? The most dominant one is the
convergence speed discussed in Section IV-B, which determines the
required number of iterations to achieve the target MSE$=10^{-4}$
(see Table   \ref{table7} and \ref{table6}). Therefore, this mainly
decides the order of the runtime curves in Fig.\ref{fig9}. The
second factor is the per-iteration runtime of the algorithms given
in Table \ref{table5}, corresponding to the number of the
matrix-vector multiplications per iteration. Therefore, we can
approximately calculate
\begin{align}
&{\text{CPU runtime}} \nonumber\\
&\approx {\text{  (\#  of iteration to Target MSE) }} \times {\text{
(Per-iteration cost).}}\nonumber
\end{align}
In some cases, the second factor highly accelerates the recovery. In
this regard, TVAMP-Condat is very competitive because it has the
smallest second factor. In all the cases of Fig.\ref{fig13}, it is
observed that TVAMP-Condat moves up its runtime from that of
TVAMP-FLSA by its cheap per-iteration cost.\footnote{Although we do
not include the NMSE convergence of TVAMP-Condat in Section IV-B, we
confirmed that TVAMP-Condat shows its convergence identical to
TVAPM-FLSA.} This also verifies our argument in Section II-A that
TVAMP can have very good scalability for large $N$ according to
choice of the numerical implementation methods of \eqref{etaTV}.

This runtime comparison  validates the low-computationality of
ssAMP-BGFD. Its fast convergence nature  and cheap per-iteration
cost provides a generally faster solution to all the cases of
Fig.\ref{fig13}. In the case (a), although ssAMP-BGFD hands over the
lead to TVAMP-Condat, it is still far better than TV-CP and EFLA,
being advantageous over TVAMP-FLSA and GrAMPA-BG. In addition, the
result of Section IV-B (see Fig.\ref{fig13}-(a)) implicates that
ssAMP-BGFD can be faster than TVAMP-Condat if the target MSE is
finer.

\section{Practical Example: Compressed Sensing Recovery of SNP Genomic Data}
\subsection{Background} In this section, we examine the ssAMP-BGFD
algorithm to the CS recovery of genomic data. In this example, we
consider a real data set of \emph{single nucleotide polymorphism}
(SNP) arrays which is a data measure for DNA copy numbers of genomic
region. The SNP data shows a 1D-PWC pattern with FD sparsity when
gene mutations occur. Specifically, the mutation causes a gene to be
either deleted from the chromosome or amplified, leading to
contiguous variation of the DNA copy numbers. Therefore, we can
identify some diseases like cancers by analyzing the SNP pattern
variation. Such a measured SNP data has been manually interpreted by
biologists, but this is time-consuming and inaccurate for two
natures of the genomic data: 1) the huge datasize and 2) severe
noise. Hence, in recent years, DSP approaches have got attention for
automatic interpretations of the genomic data, providing improved
accuracy of the analysis \cite{witten1},\cite{witten2}. The CS
framework is one line of such DSP approaches, which can resolve the
datasize problem (by measurement sampling with dimensionality
reduction) and the denoising problem (by sparsity regularization)
simultaneously.

\subsection{Experimental setup} We provide a simple demonstration
of the CS framework to the SNP data set. This data set was picked
from the chromosome 7 region of glioblastoma multiforme (GBM)
tumor\footnote{Here, we have used the SNP data set used in the work
of \cite{SNPdata},\cite{SNPdata2}.}, which has a large degree of
copy number variation. From the data set, we used the 28th, 41th,
78th, and 124th SNP samples  for this demonstration. First, we
generated  CS measurements $\underline Y$ from the noisy samples
$\underline X$ using the standard Gaussian matrix $\mathbf{H}$ with
$M/N=0.5$, then applying an algorithms to reconstruct the denoised
samples $\widehat{\underline X}$ from $\underline Y$, where we
tested some  algorithms from Table \ref{algotable}: ssAMP-BGFD
(Proposed, w/o the EM-tuning)\footnote{In the genomic applications,
the  data is severely noisy such that parameter estimation methods,
such as the EM-tuning, hardly work.}, TVAMP-Condat
\cite{TV_AMP},\cite{condat}, TV-CP \cite{CP} and GrAMPA-BG
\cite{GrAMPA}. We cannot optimally calibrate the parameters of these
tested algorithms because this example is data-driven; namely, there
are no reference signals for the recovery. Instead, we heuristically
configured each algorithm with the parameters minimizing $l_1$-norm
of the denoised sample $\widehat{\underline X}$, where we fixed
$\sigma_0=1$ and restricted the scope of the parameters to
$q\in\{10^{-6},10^{-5},10^{-4},10^{-3},10^{-2}\}, \Delta \in [0.01,
1.0], \lambda \in \{0.01, 0.1, 1, 10\}$. We set $\text{tol}=10^{-8}$
and $t^*=2000$.

\subsection{Discussion} 
\begin{figure*}[!t]
\centering
\includegraphics[width=17.5cm]{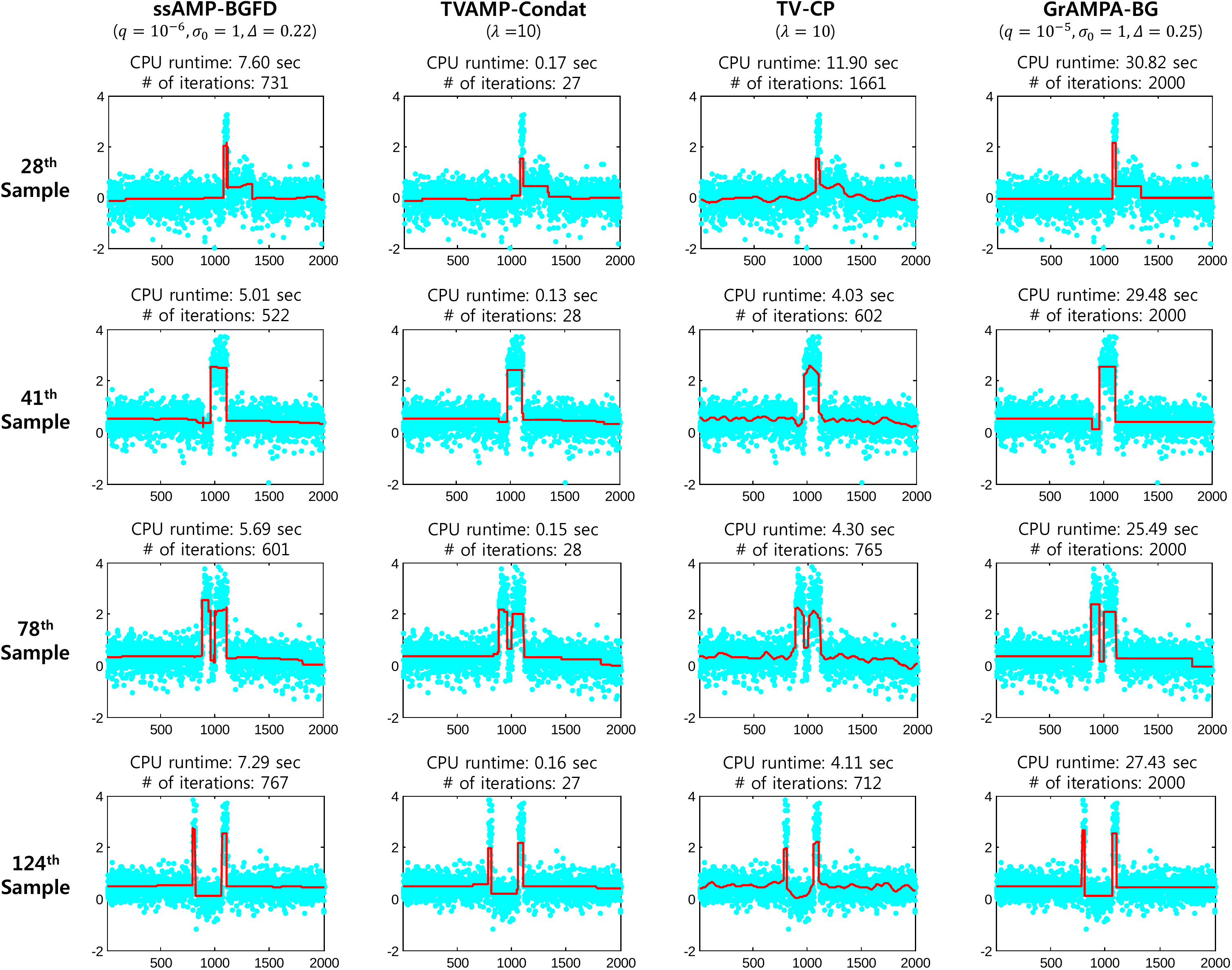}
\caption{An exemplary demonstration of the CS recovery ($M/N=0.5$)
with FD sparsity to five SNP samples of the chromosome 7 region of
GBM tumor cell. Panels in the same row are for the same SNP sample.
The blue-dots in each panel represents the original noisy SNP sample
$\underline X$. The red-solid line in each panel indicates the  SNP
sample $\widehat{\underline X}$ reconstructed and denoised from the
CS measurements $\underline Y=\mathbf{H}\underline X$ via a certain
algorithm, whose name is shown on the top of each column with its
parameter setup, where we use the standard Gaussian $\mathbf{H}$. In
addition, we provide CPU runtime (in sec) spent by each
reconstruction on the top of each panel where we used MATLAB R2013b
with Intel Core i5-750 CPU (2.67 GHz) with RAM 18GB for the CPU
runtime measuring.} \label{fig_gene}
\end{figure*}

As shown in Fig.\ref{fig_gene}, all of the
algorithms successfully recognize 1D-PWC patterns indicating the
copy number alternations in the gene samples $\widehat{\underline
X}$. TVAMP-Condat appears to be the most practical algorithm for
this SNP demonstration because of its fastest CPU runtime and its
powerful denoising ability. GrAMPA-BG has the best denoising ability
but its slow CPU runtime is demanding in practice. The TV-CP curves
do not catch the PWC shape of the SNP samples, which might cause
misidentification of copy number variations if the samples are
severely noisy. The proposed ssAMP-BGFD shows clean PWC patterns
with reasonable CPU runtime for all the SNP samples.

\section{Conclusions and Further works}
The ssAMP-BGFD algorithm, which has been proposed in the present
work, aims to solve the  CS recovery with 1D-FD sparsity  in terms
of MMSE estimation \eqref{BMMSE}. In this paper,  the algorithm
construction of ssAMP-BGFD has been mainly discussed. We have
emphasized that the key of this construction is a sum-product rule,
given in Algorithm \ref{algo1}, based on a factor graph consisting
of two types of the factor nodes: the ``s-factors" describing the
finite-difference (FD) connection of the signal $\underline X$, and
the ``m-factors" being associated with the measurement generation
\eqref{system}. From a Bayesian prospective, we have imposed a
Bernoulli-Gaussian prior \eqref{prior} on the s-factors, seeking the
FD sparsity of $\underline X$. Then, we have shown the derivation of
ssAMP-BGFD from the sum-product rule, where the Gaussian
approximation based on the central-limit-theorem and the first-order
approximation were applied to the message update with the m-factors,
and a proposed method was used to simplify the message update with
the s-factors. In addition, we have provided an EM-tuning
methodology for the prior parameter learning. The operations of
ssAMP-BGFD is fully scalar-separable and low-computational. In
addition, ssAMP-BGFD can be parameter-free with the EM-tuning,
showing phase transition closely approaching the state-of-the-art
performance by recent algorithms. We have empirically validated
these characteristics of ssAMP-BGFD compared to the algorithms
listed in Table \ref{algotable}. As a practical example, we have
applied the ssAMP-BGFD algorithm to the compressed sensing framework
with SNP genomic data set, demonstrating that ssAMP-BGFD works well
with real-world signals. An important further work is 2D/3D
extension of the ssAMP-BGFD algorithm. This work is very essential
in order to apply the ssAMP-BGFD algorithm to image denoising
applications.

\section*{Acknowledgement}
We thank Prof. Philip Schniter of Ohio State University for
providing information about the numerical settings evaluated in the
GrAMPA algorithm \cite{GrAMPA}, and  many insightful discussion
which let us to consider many details about this work. We also
express our appreciation to Prof. Hyunju Lee and her student, Jang
Ho, of  Gwangju Institute of Science and Technology  for guiding us
to handle the SNP genomic data.

\end{document}